\documentclass[onecolumn,superscriptaddress,prx,longbibliography,showpacs,preprintnumbers,amsmath,amssymb,floatfix]{revtex4-1}
\usepackage{amsmath}
\usepackage{bm}
\usepackage{graphicx}
\usepackage[ansinew]{inputenc}
\usepackage{array}
\usepackage{color}
\usepackage{amsxtra}
\usepackage{amstext}
\usepackage{amssymb}
\usepackage{latexsym}
\usepackage{dsfont}
\usepackage{braket}
\usepackage[caption=false]{subfig}
\usepackage[makeroom]{cancel}

\usepackage{dcolumn}
\usepackage{bm}
\usepackage{bbm}
\usepackage{braket}
\usepackage{mathrsfs}
\usepackage{amstext}
\usepackage{euscript}
\usepackage{txfonts}

\usepackage[unicode=true,
 bookmarks=false,
 breaklinks=false,pdfborder={0 0 1},backref=false,colorlinks=true,citecolor=blue]
 {hyperref}

\newcommand{\ms}[1]{\mbox{\scriptsize #1}}
\newcommand{\msi}[1]{\mbox{\scriptsize \textit{#1}}}

\makeatletter
\@ifundefined{textcolor}{}
{%
 \definecolor{BLACK}{gray}{0}
 \definecolor{WHITE}{gray}{1}
 \definecolor{RED}{rgb}{1,0,0}
 \definecolor{GREEN}{rgb}{0,1,0}
 \definecolor{BLUE}{rgb}{0,0,1}
 \definecolor{CYAN}{cmyk}{1,0,0,0}
 \definecolor{MAGENTA}{cmyk}{0,1,0,0}
 \definecolor{YELLOW}{cmyk}{0,0,1,0}
}

\makeatletter
\renewcommand*\env@matrix[1][*\c@MaxMatrixCols c]{%
  \hskip -\arraycolsep
  \let\@ifnextchar\new@ifnextchar
  \array{#1}}
\makeatother

\newcommand{\cref}[1]{Ref.\,\cite{#1}}

\usepackage[symbol]{footmisc}

\newcommand\footnoteref[1]{\protected@xdef\@thefnmark{\ref{#1}}\@footnotemark}

\makeatother

\begin{document}

%\preprint{}

%\title{The metrological power of nonclassical states: \\ the utility of microscopic quantum states to amplify metrological precision}

\title{The power of microscopic nonclassical states to amplify the precision of \\ macroscopic optical metrology}

\author{Wenchao Ge \footnote[1]{Correspondence and requests
for materials should be addressed to W. G.(Email: wenchaoge.tamu@gmail.com) or to K. J.(Email: dr.kurt.jacobs@gmail.com)}}
\affiliation{Department of Physics, Southern Illinois University, Carbondale, Illinois 62901, USA}
\affiliation{Institute for Quantum Science and Engineering (IQSE) and Department of Physics and Astronomy, Texas A\&M University, College Station, TX 77843-4242, USA}

\author{Kurt Jacobs \footnotemark[1]}
\affiliation{United States Army Research Laboratory, Adelphi, Maryland 20783, USA}
\affiliation{Department of Physics, University of Massachusetts at Boston, Boston, Massachusetts 02125, USA}

\author{M. Suhail Zubairy}
\affiliation{Institute for Quantum Science and Engineering (IQSE) and Department of Physics and Astronomy, Texas A\&M University, College Station, TX 77843-4242, USA}%\keywords{Keyword1, Keyword2, Keyword3}

\begin{abstract}
It is well-known that the precision of a phase measurement with a Mach-Zehnder interferometer employing strong (macroscopic) classic light can be greatly enhanced with the addition of a weak (microscopic) light field in a non-classical state. The resulting precision is much greater than that possible with either the macroscopic classical or microscopic quantum states alone. In the context of quantifying non-classicality, the amount by which a non-classical state can enhance precision in this way has been termed its ``metrological power". Given the technological difficulty of producing high-amplitude non-classical states of light, this use of non-classical light is likely to provide a technological advantage much sooner than the Heisenberg scaling employing much stronger non-classical states. To-date, the enhancement provided by  weak nonclassical states has been calculated only for specific measurement configurations. Here we are able to optimize over all measurement configurations to obtain the maximum enhancement that can be achieved by any single or multi-mode nonclassical state together with strong classical states, for local and distributed quantum metrology employing any linear or nonlinear single-mode unitary transformation. Our analysis reveals that the quantum Fisher information for \textit{quadrature displacement sensing} is the sole property that determines the maximum achievable enhancement in all of these different scenarios, providing a unified quantification of the metrological power. It also reveals that the Mach-Zehnder interferometer is an optimal network %to achieve maximum precision 
for phase sensing for an arbitrary single-mode nonlcassical input state, and how the Mach-Zehnder interferometer can be extended to make optimal use of any multi-mode nonclassical state for metrology.
\end{abstract}

\flushbottom
\maketitle
\thispagestyle{empty}

\section*{Introduction}
Measurement devices that employ quantum systems in non-classical states can outperform their classical counterparts using no more resources. The resources here are the number, $N$, of photons, phonons, spin-1/2 systems, or other elementary probe systems used by the device~\cite{giovannetti2006quantum, Caves81, PhysRevLett.71.1355, Pezze08,Liu:2013aa, Aasi:2013aa, Wollman952, Burd1163, Taylor:2016aa, PhysRevApplied.13.024037, McCuller2020, Zhao2020}. The precision of classical devices scales at most with the square root of $N$, whereas that of quantum devices can in principle scale linearly with $N$, a scaling referred to as the \textit{Heisenberg limit} \cite{PhysRevLett.71.1355}. Recent studies have tended to focus on the use of quantum systems to achieve this optimal scaling. However, as pointed out by Lang and Caves~\cite{Lang13}, since non-classical states become harder to produce the larger $N$, for measurements using light, even weak lasers will outperform the most energetic non-classical states produced to-date, something that is likely to remain true for the foreseeable future.

Even with the above limitation, non-classical states of light are remarkably powerful: few-photon non-classical states can greatly enhance the precision of measurements that employ classical states with $10^{12}$ photons~\cite{Huang84}. A well-known example is that of a Mach-Zehnder interferometer (MZI)~\cite{Lang13, Pezze08, Demkowicz15, ge2020operational, Grangier87, Li99}. A coherent state with $N_{\ms{c}}$ photons injected into one input of the MZI achieves a precision for phase measurement of $\mathcal{P} = \sqrt{N_{\ms{c}}}$~\cite{Pezze08}. (The precision is defined here as the inverse of the minimum measurement error -- see below.) Injecting a squeezed state with $N_{\ms{q}}$ photons into the second input of the MZI increases this precision to approximately $\sqrt{4\smash{N_{\ms{q}}}N_{\ms{c}}}$~\cite{Pezze08, Lang13,Demkowicz15, ge2020operational}. Thus to achieve a given precision, the addition of only 2.5 non-classical photons reduces the required power of the classical input by an order of magnitude, while 25 non-classical photons reduces it by two orders of magnitude. 

The ability of non-classical states to perform metrology is an important element in the study of non-classicality as a resource~\cite{Tan17, YadinPRX18, Kwon19, ge2020operational}. For this purpose classical resources are free, so the quantity of interest is the amount by which nonclassical states increase precision when employed with arbitrarily large coherent states, the same limit in which we are interested for practical purposes. This quantity has been termed the \textit{metrological power}, $\mathcal{M}_{\hat{\rho}}$, of a non-classical state, $\hat\rho$~\cite{Kwon19}. It provides a resource theoretic measure of nonclassicality for pure states and a witness of nonclasicality for mixed states \cite{YadinPRX18,Kwon19, ge2020operational}. 

The enhancement to otherwise classical measurements provided by weak nonclassical states has been calculated for specific scenarios such as the MZI, but the maximal enhancement enabled by \textit{any} given state (that is, the metrological power) and how to achieve it have remained lacking. Here we answer these questions for all non-classical states of light and for both local and distributed quantum metrology~\cite{PhysRevA.97.042337, Proctor18, GePRL2018,PhysRevA.97.032329, PhysRevResearch.1.032024, PhysRevResearch.2.023030}. The latter involves estimating a linear combination of a number of independent values of the same physical quantity~\cite{Guo_DQS, PhysRevX.11.031009, PhysRevLett.124.150502}. Since the quantities of interest for metrology to-date have invariably been transformations of individual modes (e.g., displacement and phase shifts) we restrict ourselves to unitary transformations of a single mode here, obtaining results for \textit{all} such transformations, including linear and nonlinear~\cite{Rivas10}, for which the metrological power is well-defined. 

%Our purpose here is to determine the maximum amount by which any non-classical state can enhance a measurement with classical light, for the regime in which the energy of the classical light is much greater than that of the non-classical state, and for metrology employing any single-mode transformation. We consider not only metrology of a single value (local metrology) but also distributed metrology  

A summary of our main results are as follows. First, we show that the quantum enhancement is always proportional to the classical precision, so that this enhancement is always an amplification of the classical precision. Second, the metrological power for the metrology of every single-mode transformation is determined by a single quantity. For  single-mode states this quantity is the quantum Fisher information (QFI) for measuring phase-space displacement using that state.  For pure states this QFI reduces to the maximum quadrature variance for the mode. For multi-mode states it is the same quantity but this time evaluated for the linear combination of the modes for which this QFI is the largest.

%This shows that the degree of squeezing completely captures the ability of weak nonclassical states to amplify the precision possible with strong laser fields. For the metrological resource theory of nonclassicality it reveals that a single resource measure will be sufficient to capture all metrology tasks for single-mode quantities~\cite{Kwon19, ge2020operational}. It also shows that in such a resource theory a state with more squeezing is more nonclassical than many states each with a little less squeezing. \comment{Maybe remove or add some supporting evidence} This parallels the result in~\cite{GePRL2018} that passive linear networks are not able to concentrate nonclassical resources located in different modes.

We show that for a single-mode nonclassical state and strong coherent state(s), the maximum precision can be obtained for single-parameter estimation merely by displacing the mode by the total available classical amplitude. For two-parameter distributed quantum metrology the balanced MZI achieves the maximum precision. For a multi-mode nonclassical state, the maximum precision can be obtained by first employing a linear network that outputs the linear combination of the modes into a single mode that has the largest QFI for displacement measurement, and then using this mode as the nonclassical input to the optimal schemes employing single-mode nonclassical states.  

Moreover, since squeezed vacuum states have the minimum energy for a given value of the quadrature variance, our results imply that a squeezed vacuum is the most energy efficient among single-mode nonclassical states for amplifying the precision for metrology of \textit{any} single-mode transformation, generalizing previous results for just phase measurements with the MZI~\cite{Lang13, ge2020operational}. 

%This paper is organized as follows. In the next section we review background information and definitions regarding local and distributed metrology and the notion of metrological power. In Section~\ref{phase} we calculate the metrological power of pure single-mode states for the metrology of phase shifts. We do this first because it allows us to illustrate our method while minimizing complexity. In Section~\ref{secall} we extend our analysis to multi-mode states and metrology using all single-mode unitary transformations for which the metrological power is well defined. In Section~\ref{sec:mixed} we extend our results to all mixed states. Section~\ref{conc} concludes. 

\section*{Results}
\subsection*{Phase metrology with a single-mode nonclassical state} 
\label{phase}
 \begin{figure}[t]
 \centering
\leavevmode\includegraphics[width = 0.7 \columnwidth]{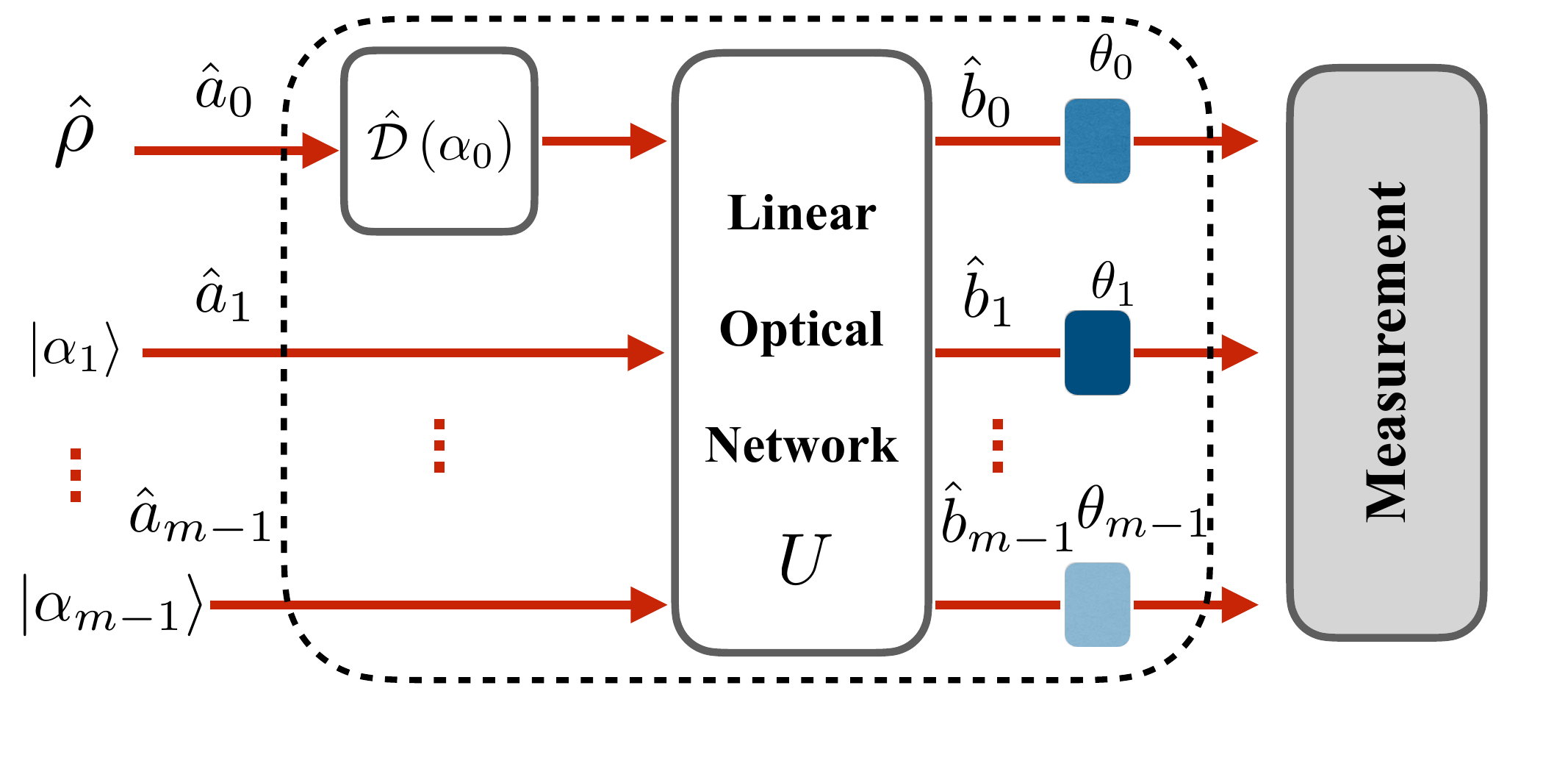}
\caption{(Color online) A distributed quantum sensing scheme. It consists of a passive linear network with $m$ input modes, one of which may contain a single-mode nonclassical state, $\hat{\rho}$, and it may be displaced by a coherent amplitude $\alpha_0$. The $m$ output modes can be sent to different locations where the respective probe transformations are applied.} 
\label{fig:scheme} 
\end{figure} 
We consider first the special case of phase sensing with a single-mode nonclassical state. This allows us to illustrate the method using the simplest nontrivial case. We generalize this method both to arbitrary single-mode transformations and multi-mode nonclassical input states in the next subsection. 

In Fig.\ref{fig:scheme} we depict a general scheme for distributed quantum metrology of $m$ phase shifts, $\theta_j$, $j=0,\ldots,m-1$ employing a nonclassical state $\hat\rho$ along with arbitrary classical resources. The mode with annihilation operator $\hat{a}_0$ contains the state $\hat{\rho}$ which we are free to displace by a coherent amplitude $\alpha_0$. Coherent states with amplitudes $\alpha_j$ are supplied in $m-1$ additional modes $\hat{a}_j\ (j=1,2,\dots,m-1)$ (we find that there is no utility in using more input modes than unknown parameters).  The modes $\hat{b}_k$ are related to the input modes $\hat{a}_j$ by the unitary $U$ so $\hat{b}_k = \sum_{j=0}^{m-1} u_{kj}\, \hat a_j$ where $u_{kj}$ are the matrix elements of $U$. Phase shift $\theta_k$ is applied to mode $\hat{b}_k$ via the transformation $\exp\left[-i\theta_k \hat{n}_k\right]$, where $\hat{n}_k \equiv \hat{b}_k^\dagger \hat{b}_k$. We will perform our analysis for the special case in which $\hat{\rho}$ is a pure state ($\hat{\rho} = \ket{\psi} \bra{\psi}$). At the end of this section, we will show how the result can be generalized to nonclassical mixed states using the result by Yu employing the convex roof~\cite{yu}. 

To evaluate the precision of multi-parameter estimation, we need the elements of the QFI matrix~\cite{toth2014quantum} for the general scheme depicted in Fig.~\ref{fig:scheme} (see Methods)
\begin{align}
    \mathcal{F}_{jk} = 4[\langle \hat{n}_j \hat{n}_k \rangle - \langle \hat{n}_j \rangle \langle \hat{n}_k \rangle], 
\end{align}
where $\braket{\cdot}$ represents the expectation value of an operator for the state $\ket{\psi}$. The complexity of this expression for $\mathcal{F}_{jk}$ comes from the fact that in terms of the input modes, $\hat{a}_j$, each of the $\hat{n}_k$ must be replaced by $\hat{b}_k^\dagger \hat{b}_k = \sum_{j=0}^{m-1} \sum_{l=0}^{m-1} u^{*}_{lk} u_{kj}\, \hat a_l^\dagger \hat a_j$.  
The key to evaluating $\mathcal{F}_{jk}$ is to employ normal ordering~\cite{SZ} of $\hat{b}_k$, noting that all but one of the input modes are in coherent states. We will denote the normal ordering of a product of mode operators in the usual way by sandwiching the product between colons. Thus $\mbox{$:\!(\hat{b}_j^\dagger \hat{b}_j)^2\!:$} = \hat{b}_j^\dagger \hat{b}_j^\dagger \hat{b}_j \hat{b}_j \not= (\hat{b}_j^\dagger \hat{b}_j)^2$. Writing $\mathcal{F}_{jk}$ in terms of normally ordered products gives 
\begin{align}
    \mathcal{F}_{jk} = \; 4 \left[ \langle :\! \hat{n}_j \hat{n}_k \!: \rangle - \, \langle \hat{n}_j \rangle \langle \hat{n}_k \rangle + \delta_{jk} \langle \hat{n}_j \rangle \right] = \; :\!\mathcal{F}_{jk}\!:\! + \,  \delta_{jk} 4 \langle \hat{n}_j \rangle 
    \label{splitqandc}
\end{align} 
where we have defined $:\!\mathcal{F}_{jk}\!:$ as four times the normally ordered covariance. Since this covariance vanishes for all classical states, and since $\langle \hat{n}_j \rangle$ (the energy of mode $j$), is effectively independent of $|\psi\rangle$ (recall that $N_{\text{q}} \ll N_{\text{c}}$), it is $:\!\mathcal{F}_{jk}\!:$ that is the  non-classical contribution to $\mathcal{F}_{jk}$. According to Eqs.~\eqref{eq:pre}, \eqref{eq:nonPre} in the section of Methods, the quantum enhancement in metrology is defined as the difference between the square precision with and without the non-classical state:
\begin{align}
    \Delta \mathcal{P}^2_{\ket{\psi}} & \leq \; \frac{\mbox{\textbf{w}}^{\ms{t}}\! :\!\mathcal{F}\! :\!\mbox{\textbf{w}}}{|\mbox{\textbf{w}}|^4} \; = \frac{1}{|\mbox{\textbf{w}}|^4}\sum_{jk} w_j w_k :\!\mathcal{F}_{jk} \! :,
    \label{eq:effectiveQFI}
\end{align} 
where $\mbox{\textbf{w}} = (w_1,\ldots,w_m)^{\ms{t}}$ are the weights in distributed phase sensing~\cite{GePRL2018} with $\sum_j|w_j|=1$.

We note next that since $\hat{b}_k$ are normally ordered, so are $\hat{a}_j$ when we make the replacement $\hat{b}_k \rightarrow \sum_j u_{kj}\hat{a}_j$. As a result, for $j \geq 1$ we can replace $\hat{a}_j$ with the coherent state amplitude $\alpha_j$ because the respective modes are in coherent states. Including the displacement of mode $\hat{a}_0$, the resulting replacement is  
\begin{align}
    \hat{b}_k & \rightarrow u_{k0} \hat{a}_0 + \sum_{j=0}^{m-1} u_{kj}\, \alpha_j = u_{k0} \hat{a}_0 + f_k 
    \label{bk}
\end{align} 
where we have defined the complex amplitudes $f_k \equiv  \sum_{j=0}^{m-1} u_{kj}\alpha_j$. 

In the limit that $|\langle a_0^n \rangle|$ and $\langle (a_0^\dagger a_0)^n \rangle$ are very much smaller than $|f_k|^n$ and $|f_k|^{2n}$ since $|f_k|\propto \sqrt{N_{\ms{c}}}\gg\sqrt{N_{\ms{q}}}$, we obtain \cite{SM}
\begin{align}
 :\!\mathcal{F}_{jk} \! : \,\, & = \, 8 |u_{k0} f_k| |u_{j0} f_j| : \! C(\hat X_{\phi_k} , \hat X_{\phi_j} ) \! : \,  
    \label{genform3}
\end{align} 
where $\phi_k = \arg[u_{k0}^* f_k]$ and $C(\hat A\hat B) \equiv \langle \hat A\hat B \rangle - \langle \hat A \rangle \langle \hat B \rangle$ as the covariance of two operators $\hat A$ and $\hat B$. Putting the expression for $:\!\mathcal{F}_{jk} \! :$ into Eq.(\ref{eq:effectiveQFI}) we can now factor the double summation to write the metrological advantage in terms of a single variance:
\begin{align} 
  \Delta \mathcal{P}^2_{\ket{\psi}}\left(N_{\text{c}},\hat{\mathbf{n}}, \mbox{\textbf{w}}\right) & \leq  \frac{8 |z|^2}{|\mbox{\textbf{w}}|^4}   : \!  V(\hat X_{\phi})\! : , \label{eq:loopy}
\end{align} 
where $V(\hat X_{\phi})\equiv C(\hat X_{\phi},\hat X_{\phi})$ and we have used the arguments $\hat{\mathbf{n}} = (\hat{n}_1,\ldots,\hat{n}_m)$ and $\mbox{\textbf{w}}$ in $\Delta \mathcal{P}^2_{\ket{\psi}}$ to denote distributed phase sensing~\cite{GePRL2018}, and $N_{\text{c}}$ as the total amount of classical resources. Here $\phi = \arg z$ and
\begin{align}
    z = \sum_k  w_k u_{k0} f_k^* . \label{eq:zform}
\end{align} 

So to obtain the metrological power we need to maximize $|z|^2$ in Eq.(\ref{eq:loopy}) over all weights $\{w_j\}$, unitary transformations $U$, and amplitudes of the classical inputs $\{\alpha_j\}$. In the supplementary materials \cite{SM}, we show that 
\begin{align} 
|z|^2\leq N_{\text{c}} |\mbox{\textbf{w}}|^4. \label{eq:ineqZ}
\end{align} 
Inserting this tight bound into the definition of the metrological power in \eqref{eq:m-power}, we obtain that for distributed phase measurement is:
\begin{align}
\label{eq:DQM}
    \mathcal{M}_{\ket{\psi}}\left(N_{\text{c}},\hat{\mathbf{n}} \right) & = 8 N_{\ms{c}} \max_\phi : \!  V(X_{\phi})\! : \; = 2N_{\text{c}} \mathcal{M}_{|\psi\rangle}^{\ms{\,F}},
\end{align} 
where $\mathcal{M}_{|\psi\rangle}^{\ms{\,F}}$ is the metrological power of $\ket{\psi}$ for quadrature displacement (force) sensing given in Eq.~\eqref{eq:MPQ}. This is one of the main results of this work that the metrological power of displacement sensing and that of phase sensing using single-mode nonclassical state are unified.

The maximum precision is given by adding to this the classical contribution to the quantum contribution Eq.~\eqref{eq:DQM}. The classical contribution is obtained according to the text above Eq.~\eqref{eq:nonPre} in the section of Methods as
\begin{align}
    \mathcal{P}_{\ms{c}}\left(N_{\text{c}},\hat{\mathbf{n}}, \textbf{w}\right) & \leq \frac{4N_c}{m|\mbox{\textbf{w}}|^2}\le4N_c,
\end{align}
where the first inequality is obtained by substituting the relation $|f_k|=|\alpha|/\sqrt{m}$ for arbitrary $\mbox{\textbf{w}}$, which is used to satisfy the lower bound in Eq.~\eqref{eq:global2}  \cite{GePRL2018}.
The last inequality is derived by invoking $m|\mbox{\textbf{w}}|^2\ge \left(\sum_j|w_j|\right)^2=1$. The maximum is achieved whenever the absolute values of all the nonzero elements of $\mbox{\textbf{w}}$ are equal. Thus both the classical and non-classical contributions in the maximum precision can be achieved simultaneously (see the supplemental material). Since for any quadrature $\hat{X}_\phi$ 
\begin{align}
   V_{|\psi\rangle}(\hat X_{\phi}) & = \; : \!  V_{|\psi\rangle}(\hat X_{\phi})\! : + \, \frac{1}{2},
\end{align}
the maximum achievable precision for phase measurement (after maximizing the passive linear network (PLN) and the weights) is thus 
\begin{align}
  \textstyle  \mathcal{P}_{|\psi\rangle} \left(N_{\text{c}},\hat{\mathbf{n}} \right) =  \textstyle  \sqrt{ \mathcal{M}_{\ket{\psi}}\left(N_{\text{c}},\hat{\mathbf{n}} \right) + \mathcal{P}_{\ms{c}}^2\left(N_{\text{c}},\hat{\mathbf{n}} \right) }  = \textstyle \sqrt{8 N_{\ms{c}} \max_\phi V_{|\psi\rangle}(\hat{X}_\phi)  } , 
  \label{loop3p}
\end{align} 
where the dependence on $\mbox{\textbf{w}}$ is dropped after the optimization procedure. This maximum precision is an amplification of the classical precision whenever $\max_\phi V_{|\psi\rangle}(\hat{X}_\phi)>1/2$. The amplification factor is 
\begin{align}
  \textstyle  \mathcal{A} = \sqrt{2\max_\phi V_{|\psi\rangle}(\hat{X}_\phi)}. 
\end{align} 

In the supplemental material \cite{SM}, we derive explicit solutions (networks) that saturate the inequality in Eq.\eqref{eq:ineqZ} and thus achieve the maximal precision. For metrology of a single parameter, the maximal precision is obtained simply by displacing the non-classical mode by the available classical energy. For two-parameter distributed quantum metrology, the maximum precision can be obtained using the balanced Mach-Zehnder interferometer. In this case 
\begin{align}
    U   =    \frac{1}{\sqrt{2}}\begin{pmatrix}
              1 & e^{i\beta} \\
             -e^{-i\beta} & 1 
             \end{pmatrix} 
\end{align}
where $\beta$ is the phase shift of the MZI beam splitter. For the maximum precision, one chooses the weightings $w_1=-w_2=1/2$.

\subsection*{Metrology of any single-mode transformation with multi-mode nonclassical states} 
\label{secall}
 \begin{figure}[t]
%\leavevmode\includegraphics[width = .75 \columnwidth]{fig2.pdf}
\centering
\leavevmode\includegraphics[width = 0.75 \columnwidth]{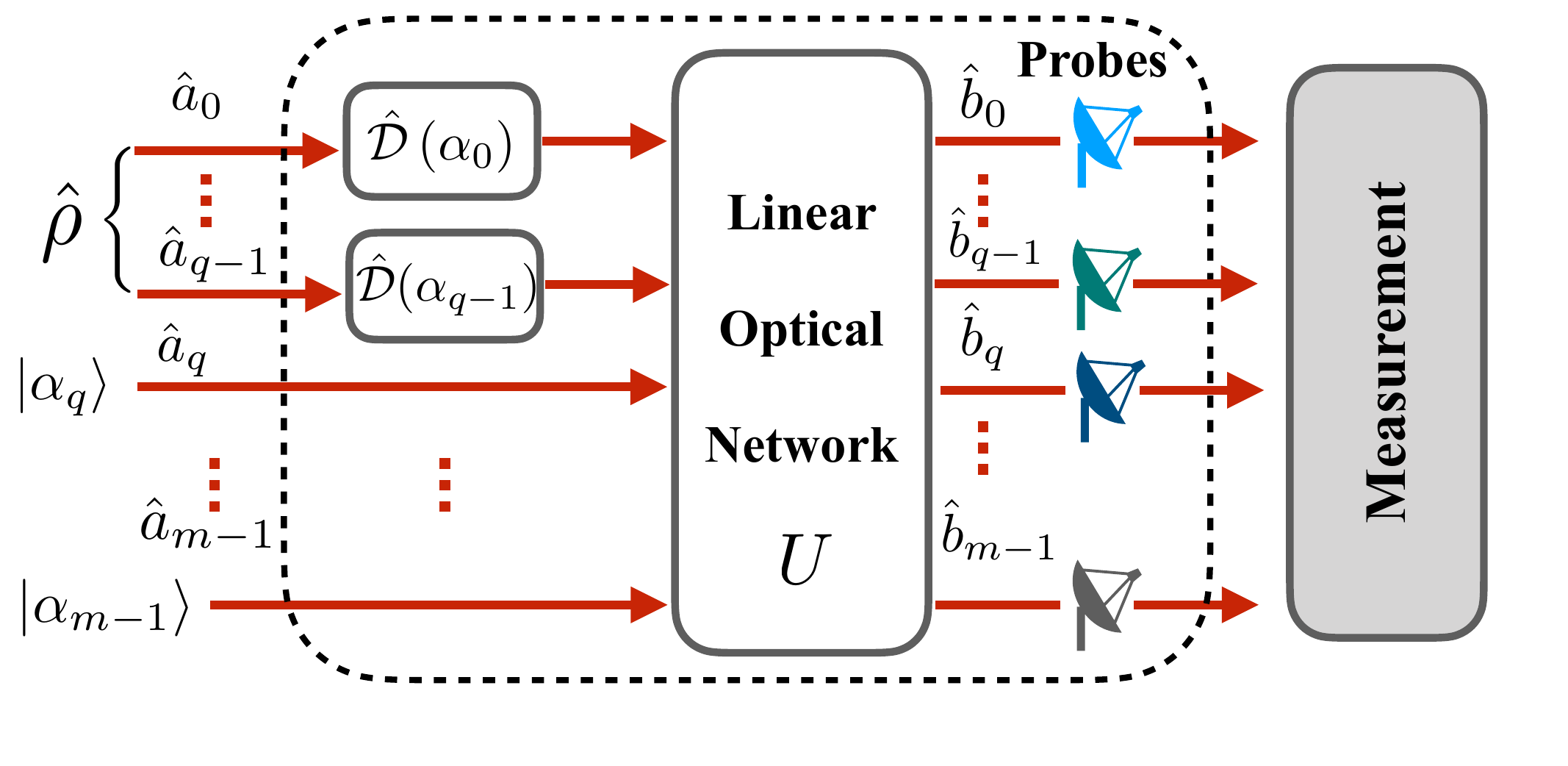}
\caption{(Color online) Distributed quantum metrology with a multimode nonclassical input state. It is a generalization to Fig.~\ref{fig:scheme} by considering a multimode nonclassical input state $\hat\rho$. The multimode nonclassical state is displaced by a multimode coherent state $\ket{\alpha_0}\otimes\cdots\ket{\alpha_{q-1}}$. Also, the probes are more general than the standard phase shifts, which are given by nonlinear operators in Eq.~\eqref{Gpkappj}} 
\label{fig:multimode} 
\end{figure} 
We now extend the results obtained in the previous section in two ways. We extend the microscopic nonclassical input to all pure multi-mode states, and we extend the phase shift to all single-mode transformations for which the metrological power is well-defined (see Fig.~\ref{fig:multimode}).    
To this end, we consider an arbitrary Hermitian single-mode transformation $\hat{G}$ expressed as a normally-ordered power series of the annihilation and creation operators:
\begin{align}
    \hat{G} & = \sum_{q=0}^{p}\sum_{j=0}^{q}\kappa_{qj} \hat{b}^{\dagger(q-j)}\hat{b}^j,
\end{align}
where $\kappa_{qj}$ are the coefficients to ensure $\hat{G}$ Hermitian. In this power series the number of mode operators in each term (the order of each term) is given by $q$. We will find that every term of order $q$ contributes to the metrological power a term proportional to $N_{\text{c}}^{q-1}$. This has two consequences. First, since the metrological power is defined in the limit of large $N_{\text{c}}$, it is undefined if the power series is infinite. We therefore restrict $\hat{G}$ to transformations for which the maximum value of $q$ is $p < \infty$. Second, in the limit of large $N_{\text{c}}$ it is only those terms of order $p$ that contribute. So for the purposes of calculating the metrological power we need only retain those terms and can thus write $\hat{G}$ as  
\begin{align}
    \hat{G} = \sum_{j=0}^p \kappa_j  \hat{b}^{\dagger (p-j)}\hat{b}^{j} .  
    \label{Gpkappj}
\end{align} 
Because $\hat{G}$ is Hermitian $\kappa_j = \kappa^{\ast}_{p-j}$, and we normalize $\hat{G}$ by setting $\sum_j \kappa_j = 1$. Subsitituting $\hat{G}$ in Eq.~\eqref{eq:QFIM}, the matrix elements of the QFI for a pure state are given by 
\begin{align}
   \mathcal{F}_{uv} = 4 \bigl[  \bigl\langle \hat{G}_{u} \hat{G}_{v} \bigr\rangle - \bigl\langle \hat{G}_{u}   \bigr\rangle  \bigl\langle  \hat{G}_{v}  \bigr\rangle \bigr] = 4\sum_{j,k}\kappa_j \kappa_k \bigl[ \bigl\langle \hat{C}_j^{(u)} \hat{C}_k^{(v)}  \bigr\rangle - \bigl\langle \hat{C}_j^{(u)} \bigr\rangle \bigl\langle \hat{C}_k^{(v)} \bigr\rangle \bigr],
\end{align} 
where $\hat{G}_{u} = \sum_{j=0}^p \kappa_j  \hat{b}_u^{\dagger (p-j)}\hat{b}_u^{j}$ and we have defined $\hat{C}_k^{(v)} = \hat{b}_v^{\dagger (p-k)}\hat{b}_v^{k}$. To split the QFI into the classical and quantum contributions we need to calculate $\bigl\langle :\! \hat{C}_j^{(u)} \hat{C}_k^{(v)} \!: \bigr\rangle$.  Deriving the identity 
 \begin{align}
    \hat b^k \hat b^{\dagger n}  & =  \hat b^{\dagger n}  \hat b^k + \sum_{j=1}^{\min(k,n)} \frac{k! n!}{(k-j)!(n-j)!j!} \hat b^{\dagger n-j} \hat b^{k-j}
\end{align}
and noting that only the terms with at least $p-2$ mode operators will contribute in the limit of large $N_{\ms{c}}$, we obtain~\cite{SM}
\begin{align}
   \mathcal{F}_{uv} = \; :\! \mathcal{F}_{uv} \! : + \, \delta_{uv}  \sum_{jk} \kappa_j \kappa_k (p-k) j \left\langle \hat{b}_u^{\dagger (2p-j-k-1)} \hat{b}_u^{j+k-1} \right\rangle  = \; :\! \mathcal{F}_{uv} \! : + \, 4\delta_{uv} \left|f_{u} \right|^{2p-2} \Biggl|  \sum_{k=1}^{p-1} \kappa_k  k  e^{2ik\gamma_{u}} \Biggr|^2 ,
   \label{Fmunuclass}
\end{align} 
where we have defined the phases $\gamma_u = \arg(f_u)$.

\subsubsection*{Classical contribution}

The classical contribution to the QFI matrix is the second term in Eq.(\ref{Fmunuclass}), namely 
\begin{align}
   \mathcal{F}_{uv}^{(\ms{c})} & \equiv  4\delta_{uv} \left|f_{u} \right|^{2p-2} \Biggl|  \sum_{k=1}^{p-1} \kappa_k  k  e^{2ik\gamma_{u}} \Biggr|^2. 
   \label{Fmunuclass3}
\end{align} 
We now recall that to maximize the precision, the QFI matrix must be such as to have $\mbox{\textbf{w}}$ as an eigenvector \cite{GePRL2018}. Since we see from Eq.(\ref{Fmunuclass3}) that the classical contribution to this matrix, $\mathcal{F}_{uv}$, is diagonal, this can only be satisfied if $\mathcal{F}$ is proportional to the identity or $\mbox{\textbf{w}}$ has only one non-zero element. In fact, the latter is merely the special case of the former in which $m=1$. We also note that the final summation in Eq.(\ref{Fmunuclass}) is maximized by choosing the appropriate value for the phases $\gamma_u$. Since these phases can be chosen arbitrarily and independently of $|f_u|$, the classical contribution to the QFI matrix is  
\begin{align}
    \mathcal{F}_{uv}^{(\ms{c})} & = 4\delta_{uv} B(\bm{\kappa}) |f_u|^{2(p-1)} = 4\delta_{uv} B(\bm{\kappa}) \frac{N_{\ms{c}}^{p-1}}{m^{p-1}}, 
    \label{Fmunuclass2}
\end{align}
where 
\begin{align}
    B(\bm{\kappa}) \equiv  \max_\gamma\Biggl|  \sum_{k=1}^{p-1} \kappa_k  k  e^{2ik\gamma} \Biggr|^2 . 
    \label{kappaj}
\end{align}
To obtain the RHS of Eq.(\ref{Fmunuclass2}), we have chosen all the $|f_u|$ to be equal ($|f_u|^2 = N_{\ms{c}}/m$) so that $\mathcal{F}_{uv}$ is proportional to the identity. According to the section of Methods, the resulting precision is now 
\begin{align}
    \mathcal{P}_{\ms{c}}^2\left(N_{\text{c}},\hat{\mathbf{G}}, \mbox{\textbf{w}}\right) & \le  4 B(\bm{\kappa}) \frac{N_{\ms{c}}^{p-1}}{m^{p-1}} \frac{1}{|\mbox{\textbf{w}}|^2} \le 4 B(\bm{\kappa}) \frac{N_{\ms{c}}^{p-1}}{m^{p-2}}\le 4 B(\bm{\kappa}) N_{\ms{c}}^{p-1} \label{maxP2cnonl},
\end{align}
we have used the arguments $\hat{\mathbf{G}} = (\hat{G}_1,\ldots,\hat{G}_m)$ and $\mbox{\textbf{w}}$ in $\mathcal{P}_{\ms{c}}^2$ to denote the general distributed sensing with an arbitrary single-mode Hermitian transformation. Here the second inequality is maximized by choosing $\mbox{\textbf{w}}$ to be equally distributed over the $m$ parameters. Note that when $p=2$ (phase measurement) the maximum precision is the same for any number of parameters, i.e., independent of $m$. However, when the transformation $\hat G$ is nonlinear $(p \geq 3)$ the precision reduces as the number of parameters is increased. Thus the maximum classical precision is obtained for $m=1$ as given in the last inequality in the above equation.

% The classical contribution to the square precision for the global estimation in Eq.~\eqref{eq:global0} is thus 
% \begin{align}
%     \mathcal{P}_{\ms{c}}^2\left(N_{\text{c}},\hat{\mathbf{G}}^{(m)}, \mbox{\textbf{w}}\right) &=\frac{4}{\sum_u \frac{w_u^2}{v_u}}\le \frac{4\sum_u v_u}{(\sum_u |w_u|)^2}=4\sum_u v_u
% \end{align}
% where $v_u=  \left|f_{u} \right|^{2(p-1)} \Biggl|  \sum_{k=1}^{p-1} \kappa_k  k  e^{2ik\gamma_{u}} \Biggr|^2$ and $\hat{\mathbf{G}}^{(m)} = (\hat{G}^{(0)}, \ldots, \hat{G}^{(m-1)})$ denotes the vector of the nonlinear transformations. The inequality becomes equality when $\mbox{\textbf{w}}$ and $\mbox{\textbf{v}}$ are linearly dependent. The factor $\Biggl|  \sum_{k=1}^{p-1} \kappa_k  k  e^{2ik\gamma_{u}} \Biggr|^2$ can be maximized by choosing a suitable linear network and input coherent states, and thus we define $B(\bm{\kappa}) \equiv  \max_\gamma\Biggl|  \sum_{k=1}^{p-1} \kappa_k  k  e^{2ik\gamma} \Biggr|^2$ for all $v_u$ by omitting $u$ in $\gamma_u$. Since $\sum_u |f_u|^2=N_c$ and $p-1>1$, $\sum_u |f_u|^{2(p-1)}\le N_{\ms{c}}^{p-1}$ when the distributed multi-parameter estimation scheme is reduced to a single-parameter estimation. This is due to the nonlinear nature of the generator. The resulting  maximum classical square precision is 
% \begin{align}
%     \mathcal{P}_{\ms{c}}^2\left(N_{\text{c}},\hat{\mathbf{G}}^{(m)}\right) &  = 4 N_{\ms{c}}^{p-1}  B(\bm{\kappa})
% \end{align} 

% ---------------------------------------------

\subsubsection*{Quantum contribution}

We now turn to calculating the quantum contribution to the QFI, namely $:\! \mathcal{F}_{uv} \! :$. Instead of a single input mode containing a nonclassical state, now input modes with mode operators $a_0, a_1, \ldots, a_{q-1}$ contain a joint nonclassical state (Fig.~\ref{fig:multimode}). As a result the transformation in Eq.(\ref{bk}) is replaced by  
\begin{align}
    \hat{b}_u \rightarrow \sum_{l=0}^{q-1} u_{ul}\hat{a}_l + f_u
\end{align}
and as before $f_{u} = \sum_{l=0}^{m-1} u_{ul} \alpha_{l}$. We rewrite this sum over the nonclassical modes $a_l$ as a single mode operator: 
\begin{align}
    \hat{b}_u \rightarrow g_u \hat{d}_u + f_u
\end{align}
where $\hat{d}_u = \frac{1}{g_u} \sum_{l=0}^{q-1} u_{ul} \hat a_l$ and $g_u = \left( \sum_{l=0}^{q-1} |u_{ul}|^2 \right)^{\! 1/2} $.
Making the above replacement in $:\! \mathcal{F}_{uv} \! :$ and keeping only those terms with no more than two factors of $\hat{d}_u$ as in the previous section, we find that we can write it in the form~\cite{SM}
\begin{align}
  \lim_{N_{\ms{c}} \rightarrow \infty}  :\! \mathcal{F}_{uv} \!:  \; = 4\left(\braket{:\!\hat{A}_u\hat{A}_v\!:}-\braket{\hat A_u}\braket{\hat{A}_v}\right),
    \label{eq:Fuv}
\end{align}
where     $\hat{A}_u = \sum_{j=0}^p \kappa_j f_u^{\ast p-j-1}f_u^{j-1} g_{u}\left((p-j)f_u \hat{d}_u^{\dagger}+jf_u^{\ast}\hat{d}_u\right).$ By substituting $:\! \mathcal{F}_{uv} \!:$ in Eq.~\eqref{eq:effectiveQFI}, we obtain
\begin{align}
     \Delta \mathcal{P}^2_{\ket{\psi}}\left(N_{\text{c}},\hat{\mathbf{G}}, \mbox{\textbf{w}}\right) \le \max_{\text{PLN}}\frac{8|z|^2}{|\mbox{\textbf{w}}|^4} \!:\!V_{\ket{\psi}}\left( \hat{X}_d \right)\!: . 
     \label{P2nonl} 
\end{align}
Here $|z|^2 = B(\bm{\kappa}) \sum_{l=0}^{q-1} |c_l|^2$ and $\hat{X}_d = (\hat{d} + \hat{d}^\dagger)/\sqrt{2}$, where the mode operator $\hat{d}$ is the linear combination of nonclassical input modes 
\begin{align}
    \hat{d} & = \frac{ \sum_{l=0}^{q-1 }c_l \, \hat{a}_l}{\sqrt{ \sum_{l=0}^{q-1} |c_l|^2 }} 
\end{align}
with coefficients $c_l = \sum_u w_u |f_u|^{p-1}u_{ul} = \mathbf{v}\cdot\mathbf{u}_l $. Here the elements of the vector $\mathbf{v}$ are $v_u = w_u |f_u|^{p-1}$.  We need to choose $\hat{d}$ so that it is the linear combination of the non-classical input modes that has the maximum quadrature variance, and also maximize the sum $\sum_{l=0}^{q-1}|c_l|^2$. To this end, we first note that the set of $q$ vectors $\mathbf{u}_l$ are orthonormal ($q\leq m$) and can be chosen arbitrarily. We choose them so that the vector $\mbox{\textbf{v}}$ lies in the $q$-dimensional space that they span. This automatically ensures that
\begin{align}
    \sum_{l=0}^{q-1} |c_l|^2 = |\mathbf{v}|^2 . 
\end{align}
which is its maximum value. Since the vectors $\mbox{\textbf{u}}_l$ are a basis for a $q$-dimensional space containing $\mbox{\textbf{v}}$, we can choose the coefficients $c_l$ arbitrarily merely by rotating this basis. In particular we can choose $c_l$ so that $\hat{d}$ gives the maximum quadrature variance. 

We now use the fact that $|\mbox{\textbf{v}}|^2 = (\mbox{\textbf{v}}\cdot \mbox{\textbf{r}})^2$ when $\mbox{\textbf{r}}$ is a unit vector aligned with $\mbox{\textbf{v}}$. This allows us to write 
\begin{align}
  |z|^2  = B(\bm{\kappa})  |\mbox{\textbf{v}}|^2 = B(\bm{\kappa}) \, \Biggl( \sum_{u=0}^{m-1} w_u |f_u|^{(p-1)} r_u \Biggr)^2 . 
\end{align}
Apart from the fact that $f_u$ is raised to a higher power, the summation in $|z|^2$ has exactly the same form as that in Eq.~\eqref{eq:zform} for phase sensing with a single-mode nonclassical state. We therefore use the same procedure~\cite{SM} to optimize it and obtain  
\begin{align}
     |z|^2 \leq  N_{\ms{c}}^{p-1} B(\bm{\kappa}) |\mbox{\textbf{w}}|^4 . 
     \label{zmaxnonl}
\end{align}
This time, however, the upper bound can only be saturated when the vector $(f_0,f_1,\dots,f_{m-1})^T$ has only one non-zero element. Since $\sum_u |f_u|^2=N_c$ we have $\sum_u|f_u|^{2(p-1)}\le N_c^{p-1}$ where for $p\geq 3$ the equality holds only if the vector has only one non-zero element. In this case, to optimize the precision, $\mbox{\textbf{w}}$ must also have only one non-zero element (single-parameter metrology). Thus for a fixed available classical energy, for nonlinear metrology ($p\ge 3$) the precision decreases as the number of measured parameters is increased. This is not the case for linear metrology (phase measurement). 

Substituting Eq.(\ref{zmaxnonl}) into Eq.(\ref{P2nonl}), we obtain the maximum nonclassical contribution to the square precision, which is also the metroloical power:  
\begin{align}
    \mathcal{M}_{\ket{\psi}}\left(N_{\text{c}},\hat{\mathbf{G}}\right) = \max_{\mbox{\text{PLN},\textbf{w}}} \Delta \mathcal{P}^2_{\ket{\psi}}\left(N_{\text{c}},\hat{\mathbf{G}}, \mbox{\textbf{w}}\right) = 8  N_{\ms{c}}^{p-1} B(\bm{\kappa}) \max_{d}   \!:\!V_{\ket{\psi}}\left( \hat{X}_d \right)\!: ,
    \label{maxP2qnonl}
\end{align}
where the maximization over $d$ is over all passive linear transformations of the non-classical input modes. Adding together the classical and nonclassical contributions to the precision, Eqs.(\ref{maxP2cnonl}) and (\ref{maxP2qnonl}), gives us the maximum precision obtainable with nonlinear metrology, strong coherent states with total photon number $N_{\ms{c}}$, and weak non-classical states: 
\begin{align}
       \mathcal{P}_{\ket{\psi}}\left(N_{\text{c}},\hat{\mathbf{G}} \right) & = \sqrt{  8N_{\ms{c}}^{p-1} B(\bm{\kappa}) \max_{d} V\left( \hat{X}_d \right) }.
   \label{PKp}
\end{align}
The expression for the maximum precision, Eq.(\ref{PKp}), shows that for a multimode input this precision is the same as for a single-mode input when the single mode has the same maximum quadrature variance as the optimal linear combination of the multiple nonclassical input modes. This shows us immediately that the following two-stage linear network will achieve maximal precision. First we construct a network that takes the multi-mode noclassical state as an input and produces the linear combination of the input modes that has the maximal quadrature variance at one of its outputs. Second, we feed this output into a network in the scheme that achieves the maximum precision for a single-mode non-classical input. We show in the supplemental material \cite{SM} that the MZI is one such network.  

\subsection*{Metrological power for mixed states\label{sec:mixed}}

So far we have calculated the metrological power only for pure states. We can now use Yu's theorem~\cite{Toth13,yu}, given in Eq.(\ref{conroof}), to extend our results to all mixed states. Yu's theorem is only valid for single-parameter metrology, but since the maximum non-classical contribution to the square precision (the metrological power) can be achieved for $m=1$, Yu's theorem is sufficient for our purposes. We have
\begin{align} 
   \mathcal{F}\left(\hat{\rho},\hat{G}\right) 
      & =  \min_{\{p_n,\ket{\psi_n}\}} \sum_n p_n \mathcal{F} \left(\ket{\psi_n}\bra{\psi_n},\hat{G}\right) \nonumber \\
      & =  8 N_{\ms{c}}^{p-1} B(\bm{\kappa}) \max_{d} \left[ \min_{\{p_n,\ket{\psi_n}\}}  \sum_n p_n V_{\ket{\psi_n}}\left( \hat{X}_d \right) \right]  \nonumber \\ 
      & =  8 N_{\ms{c}}^{p-1} B(\bm{\kappa}) \max_{d} \mathcal{F}\left(\hat{\rho},\hat{X}_d\right) 
      \label{mixed} 
\end{align} 
where the second line is obtained by using the fact that for single-parameter metrology the QFI is the square precision which for pure states is in turn given by Eq.(\ref{PKp}). We recall that $B(\bm{\kappa})$ is defined in Eq.(\ref{kappaj}), and the parameters ${\kappa_j}$ define the transformation $\hat G$, given in Eq.(\ref{Gpkappj}). Here $\mathcal{F}\left(\hat{\rho},\hat{X}_d\right)$ is the QFI for quadrature displacement of the operator $\hat{X}_d = (\hat{d} + \hat{d}^\dagger)/\sqrt{2}$. The maximization is over a mode $\hat{d}$ that is a passive linear transformation of the non-classical input modes. 
Note that there is only a single maximization over the mode $\hat{d}$ because when using the mixed state for metrology we can only use one linear network (we cannot use a different linear network for each state in the decomposition of $\hat\rho$). We note that there is a closed-form expression for the QFI, which can be used to calculate $\mathcal{F}\left(\hat{\rho},\hat{X}_d\right)$ for a given $\hat\rho$ and $\hat X_d$~\cite{Braunstein:1994aa}.  

Performing the analysis in Eq.(\ref{mixed}), but splitting the QFI into its quantum and classical parts, we have the equivalent result for the metrological power for a mixed state
\begin{align}
   \mathcal{M}_{\rho}\left(N_{\text{c}},\hat G\right)   = 8 N_{\ms{c}}^{p-1} B(\bm{\kappa}) \max_{d} \left[ \min_{\{p_n,\ket{\psi_n}\}}  \sum_n p_n :\! V_{\ket{\psi_n}}\left( \hat{X}_d \right)\!:  \right] = 4 N_{\ms{c}}^{p-1} B(\bm{\kappa}) \mathcal{M}_{\rho}^{\ms{\,F}} .
\end{align}

\section*{Discussion}

\label{conc}

Weak non-classical light is able to amplify the precision of measurements that employ much stronger coherent light. Here we have determined the maximum amplification that can be achieved in this way by every quantum state of light (more generally, any bosonic field), for local and distributed quantum metrology that employs any single-mode transformation. We have shown that for the measurement of all single-mode quantities the maximum amplification depends on the same quantity, being the QFI for displacement metrology. For single-mode pure states this QFI is simply the maximum quadrature variance. Our analysis also reveals the linear networks that can be used to achieve the maximum precision. 

The method we have used to obtain our results should also enable answering the same question for multi-mode transformations, which constitutes an interesting question for future work. We also expect that our results will have applications to the resource theory of nonclassicality, something that we have not explored here.

\section*{Methods}

\subsection*{Local and distributed quantum metrology} 
\label{secdef}

Quantum metrology refers to the measurement of a classical quantity or quantities using a quantum system as a probe. First consider metrology of a single quantity, $\xi$. For a quantum system to act as a probe its state must be affected in some way by $\xi$ so that information about $\xi$ can be extracted by making a measurement on the system. We can write the effect of $\xi$ on the system as the action of an operator $U_\xi = e^{-i\xi \hat{G}}$ where $\hat{G}$ (the ``metrological transformation") is a Hermitian operator~\cite{giovannetti2006quantum}. As an example, if we are measuring the size of a force then $\hat{G} = \hat{b} e^{-i\phi}/\sqrt{2} + \mbox{H.c}\equiv\hat{X}_{\phi}$ is the position operator, where $\hat b$ is the mode operator at the probe and $\phi$ is the quadrature phase. If we are measuring a phase shift induced in a mode by a distance or a duration, then $\hat G = \hat b^\dagger \hat b$. The amount of information that can be obtained  about $\xi$ by measuring the system depends on the state in which the system is prepared prior to the action of $U(\xi)$. If we denote this state by $\hat\sigma$, then the maximum information that can be obtained about $\xi$ (the ``sensitivity" of the probe) is captured by the QFI, denoted by $\mathcal{F}(\hat{\sigma},\hat{G})$. For a pure state $\hat{\sigma}=\ket{\psi}\bra{\psi}$, the QFI reduces to four times the variance of $\hat{G}$~\cite{toth2014quantum}, namely 
\begin{align}
   \mathcal{F}(\ket{\psi_n}\bra{\psi_n},\hat{G}) \equiv 4V_{|\psi\rangle}(\hat{G}) = 4 \left( \langle \hat{G}^2 \rangle - \langle \hat{G} \rangle^2 \right) , 
\end{align}
where $\langle \hat{A} \rangle \equiv \langle\psi | \hat{A} | \psi \rangle$ for any operator $\hat{A}$. For mixed states, Yu's theorem states that the QFI can be written in terms of that for pure states as~\cite{yu,Toth13} 
\begin{align} 
\mathcal{F}\left(\hat{\sigma},\hat{G}\right) & =  \min_{\{p_n,\ket{\psi_n}\}} \sum_n p_n \mathcal{F} \left(\ket{\psi_n}\bra{\psi_n},\hat{G}\right)  , 
  \label{conroof}
\end{align} 
where the minimization is over all ensembles that decompose $\hat{\sigma}$ (all ensembles $\{p_n,\ket{\psi_n}\}$ for which $\hat{\sigma} = \sum_n p_n\ket{\psi_n}\bra{\psi_n}$ and $p_n>0, \; \forall n$). The expression on the RHS of Eq.(\ref{conroof}) is referred to as the \textit{convex roof} of $\mathcal{F}$.

The QFI quantifies the minimum error with which the parameter $\xi$ can be obtained, $\Delta \xi$, by measuring the probe system given the initial state $\hat{\sigma}$ and transformation $\hat G$ via the relation~\cite{helstrom1969quantum, Braunstein:1994aa}
\begin{align}
    \Delta \xi \ge \frac{1}{\sqrt{M \mathcal{F}\left(\hat{\sigma},\hat{G}\right)}}
\end{align}
where $M$ is the (sufficiently large) number of repetitions of the metrology procedure. This relationship is referred to as the quantum Cram\'er-Rao bound~\cite{helstrom1969quantum, Braunstein:1994aa}.

Defining the \textit{precision} of a metrology protocol, $\mathcal{P}\left(\hat{\sigma},\hat{G}\right)$, as the inverse of the minimum error per root repetition, we have 
\begin{align}
\label{eq:precision}
    \mathcal{P}\left(\hat{\sigma},\hat{G}\right) \equiv\sqrt{\mathcal{F}\left(\hat{\sigma},\hat{G}\right)} . 
\end{align}

Distributed quantum metrology is a straightforward generalization of the metrology process considered above in which a quantum system is used to measure a function of a set of $m$ parameters $\{\xi_j\}$~\cite{PhysRevA.97.032329,PhysRevA.97.042337,GePRL2018,Proctor18, Qian2019}. Usually these parameters are considered to be values of the same physical quantity at different locations. Here we will restrict ourselves to determining (estimating) a linear combination of the parameters. Our analysis is also applicable to the simultaneous estimation of all the parameters~\cite{PhysRevLett.111.070403}. 

In Fig.~\ref{fig:scheme}, we display a distributed quantum sensing scheme with an initial input state $\hat{\sigma}=\mathcal{D}\left(\alpha_0\right)\hat\rho\mathcal{D}^{\dagger}\left(\alpha_0\right)\otimes \ket{\alpha_1}\bra{\alpha_1}\cdots\otimes\ket{\alpha_{m-1}}\bra{\alpha_{m-1}}$, in which one of the input modes contains an arbitrary quantum state, $\hat\rho$, displaced by a coherent displacement $\mathcal{D}\left(\alpha_0\right)\equiv \exp(\alpha_0 a_0^{\dagger}-\alpha_0^{\ast}a_0)$ and the rest are in coherent states, $\ket{\alpha_j} \ (j=1,\ 2,\ \cdots, m-1)$. After combining all the input modes with an arbitrary passive linear network each transformation is applied to a different mode. This allows each mode to be sent to a different location where the respective transformations are applied. Distributed quantum metrology is not the most general way to implement a measurement of a global parameter. In the supplemental materials~\cite{SM}, we depict the most general linear scheme for this purpose. We show that in the regime in which we are interested here, that is when the classical input energy is much larger that that of the nonclassical states, the most general scheme can improve upon the precision of distributed quantum metrology \textit{only} by the factor $m$, the origin of which is a trivial \textit{classical} effect resulting from applying all the phase shifts in sequence to a single mode.

As per Fig.~\ref{fig:scheme}, defining mode operators $\hat{b}_j, j=0,\ldots,m-1$ for each of $m$ modes at the probe, and operators $\hat{G}_j = f(\hat{b}_j,\hat{b}_j^\dagger)$ in which $f$ is some function, the action of the set of parameters on the probe system is given by the product
\begin{align}
    U_{\boldsymbol{\xi}} = \prod_j \exp[-i\xi_j \hat{G}_j].
\end{align} 
If the input state $\hat\rho$ is a pure state then the QFI for this multi-parameter estimation problem, which we again denote by $\mathcal{F}$, is now a matrix whose elements are \cite{helstrom1969quantum, toth2014quantum}  
\begin{align}
\label{eq:QFIM}
    \mathcal{F}_{jk} = 4 \left( \langle \hat{G}_j \hat{G}_k \rangle -\langle  \hat{G}_j \rangle \langle  \hat{G}_k \rangle \right) . 
\end{align}
For simultaneous estimation of all the parameters $\xi_j$ with a set of unbiased estimators $\Xi_j$, the error of the estimators is now given by a covariance matrix, $\text{cov}\left(\boldsymbol{\Xi}\right)$, whose matrix elements are 
\begin{align}
    \text{cov}\left(\boldsymbol{\Xi}\right)_{jk} = \braket{(\Xi_j-\xi_j)(\Xi_k-\xi_k)}, 
\end{align}
where $\braket{O}$ is the average value of the quantity $O$. Note that here the average is not only a quantum expectation over the joint multi-mode state input to the passive linear network but also over the possible values of the parameters and the results of the measurement for each repetition of the metrology scheme. The multi-parameter version of the quantum Cram\'er-Rao bound~\cite{helstrom1969quantum,toth2014quantum} is 
\begin{align}
    \text{cov}\left(\boldsymbol{\Xi}\right) \geq (M\mathcal{F})^{-1}, 
\end{align}
where $M$ is again the number of independent repetitions of the metrology process. 

For estimating a linear combination of the parameters, namely 
\begin{align}
    \xi = \sum_j w_j \xi_j, \;\;\;\; \mbox{with} \;\; \sum_j |w_j| = 1, 
    \label{eq:global0}
\end{align}
which is referred to as a global estimate \cite{PhysRevA.97.032329,PhysRevA.97.042337,GePRL2018,Proctor18, Qian2019}, the measurement uncertainty in the estimate of $\xi$ is~\cite{GePRL2018} 
\begin{align}
\label{eq:global1}
    \Delta\xi \ge \sqrt{\frac{\mbox{\textbf{w}}^{\ms{t}}\mathcal{F}^{-1}\mbox{\textbf{w}}  }{ M } } 
\end{align} 
in which we have defined the vector $\mbox{\textbf{w}} = (w_1,\ldots,w_m)^{\ms{t}}$, and $|\mbox{\textbf{w}}|^2 = \mbox{\textbf{w}}^{\ms{t}}\mbox{\textbf{w}} = \sum_j w_j^2$. 
The above inequality is not well-suited to minimizing $\Delta\xi$, however, because it involves the inverse of the QFI matrix. We can obtain a much simpler lower bound by using the Cauchy-Schwartz inequality to show that~\cite{GePRL2018}
\begin{align}
\label{eq:global2}
    \Delta\xi \ge \sqrt{\frac{\mbox{\textbf{w}}^{\ms{t}}\mathcal{F}^{-1}\mbox{\textbf{w}}  }{ M } } \ge  \frac{|\mbox{\textbf{w}}|^4}{\sqrt{M\mbox{\textbf{w}}^{\ms{t}}\mathcal{F}\mbox{\textbf{w}} }} 
\end{align}
The beauty of the second inequality is that it becomes an equality when $\mbox{\textbf{w}}$ is an eigenvector of the QFI matrix. We can therefore find the maximum precision by minimising the RHS of the above chain of inequalities under the constraint that $\mbox{\textbf{w}}$ is an eigenvector of $\mathcal{F}$. 

We define the precision of a distributed quantum metrology protocol by 
\begin{align}
\label{eq:pre}
    \mathcal{P}\left(\hat{\sigma},\hat{\mathbf{G}},\mathbf{w}\right) \equiv  \left( \mbox{\textbf{w}}^{\ms{t}}\mathcal{F}^{-1}\left(\hat{\sigma},\hat{\mathbf{G}}\right)\mbox{\textbf{w}} \right)^{-1/2}  \leq  \frac{1}{|\mathbf{w}|^2} \sqrt{\mbox{\textbf{w}}^{\ms{t}}\mathcal{F}\left(\hat{\sigma},\hat{\mathbf{G}}\right)\mbox{\textbf{w}}} , 
\end{align} 
where $\hat{\mathbf{G}} = (\hat{G}_1,\ldots,\hat{G}_m)$ denotes the set of transformations for $m$ probes. Note that a distributed quantum metrology protocol automatically reduces to a single-parameter protocol when $m=1$.

\subsection*{Metrological power}

A classical metrology procedure is one in which the initial state, $\hat\sigma$, consists of any number of modes in coherent states $|\alpha_k\rangle$. An optimal classical metrology procedure for a given set of transformations $\{\hat{G}_j\}$ is one that achieves the maximum QFI for the total average number of photons, $N_{\ms{c}} = \sum_k |\alpha_k|^2$. Obtaining an optimal protocol, by definition, involves maximizing over all passive linear networks as well as the choice of weightings, $\mathbf{w}$. 

Consider an optimal classical metrology procedure $\mathbb{M}$, that employs $m$ transformations $\hat{\mathbf{G}}$ to measure a single ($m=1$) or global parameter $\xi = \sum_{j=0}^{m-1} w_j \xi_j$. Let us denote the precision of $\mathbb{M}$ by $\mathcal{P}_{\ms{c}}(N_{\ms{c}},\hat{\mathbf{G}},\mathbf{w})$. Now define an optimal metrology protocol that employs classical states with the same total number of photons $N_{\ms{c}}$ and transformations $\hat{\mathbf{G}}$, but this time with an additional mode containing a non-classical state $\hat\rho$ with average photon number $N_{\ms{q}}$. Since $N_{\ms{q}}$ is required to be negligible compared to $N_{\ms{c}}$, there is little point in including $N_{\ms{q}}$ in the resource count for the protocol. Hence, we will denote the precision of this protocol by $\mathcal{P}_{\hat\rho}(N_{\ms{c}},\hat{\mathbf{G}},\mathbf{w})$. The increase in the square of the precision resulting from adding the nonclassical state is 
\begin{align}\label{eq:nonPre}
    \Delta \mathcal{P}^2_{\hat\rho}(N_{\ms{c}},\hat{\mathbf{G}}, \mathbf{w})  = \mathcal{P}_{\hat\rho}^2(N_{\ms{c}},\hat{\mathbf{G}},\mathbf{w}) - \mathcal{P}_{\ms{c}}^2(N_{\ms{c}},\hat{\mathbf{G}},\mathbf{w}). 
\end{align}
We wish to define the \textit{metrological power} as this increase in the square precision as $N_{\ms{c}}\rightarrow\infty$ for a fixed $N_{\ms{q}}$. We use the square of the precision so that the metrological power is linear in the QFI. If $\Delta\mathcal{P}^2_{\hat\rho}(N_{\ms{c}},\hat{\mathbf{G}},\mathbf{w})$ depends on $\mathbf{w}$ and the PLN involving the coherent states and the unitary $U$, then we also maximize over $\mathbf{w}$ and the PLN. To be precise, in our definition we have to be a little careful because the metrological power scales with the resource $N_{\ms{c}}$ and thus tends to infinity as $N_{\ms{c}} \rightarrow\infty$. 

Allowing the elements of $\hat{\mathbf{G}}$ to take the general form 
\begin{align}
    \hat G_j = \sum_{l=0}^p \sum_{r=0}^l \kappa_{lr} \hat b_j^{\dagger l-r} \hat b_j^{r},  
\end{align}
we will find that for large $N_{\ms{c}}$ this maximum increase is proportional to $N_{\ms{c}}^{p-1}$. We can thus define the metrological power as 
\begin{align}
\label{eq:m-power}
    \mathcal{M}_{\hat{\rho}}(N_{\ms{c}},\hat{\mathbf{G}})  = \max_{\mathbf{w},\text{PLN}} \left[ \lim_{N_{\ms{c}}\rightarrow\infty} \frac{\Delta \mathcal{P}^2_{\hat\rho}(N_{\ms{c}},\hat{\mathbf{G}},\mathbf{w}) }{N_{\ms{c}}^{p-1}}\right] N_{\ms{c}}^{p-1} . 
\end{align} 

There is one transformation for which the maximum precision is straightforward to calculate and thus the metrological power is already known~\cite{Kwon19}. This transformation is a displacement in phase space (corresponding to measuring force or acceleration), for which $\hat{G} = \hat{X}_{\phi} \equiv \hat{b} e^{-i\phi}/\sqrt{2} + \mbox{H.c}$. This transformation is unique in that the maximum precision is achieved using the nonclassical state alone; there is no benefit to combining the nonclassical state with classical states. Thus the QFI for a displacement transformation using state $|\psi\rangle$, which we will denote by $\mathcal{F}_{\phi}$, is given simply by four times the variance of $\hat{X}_{\phi}$ for $|\psi\rangle$. Since the quadrature angle, $\phi$, can be selected merely by using a phase shift, we must maximize over it to determine the maximum precision~\cite{Knobel:2003aa, Hoff:2013aa}. According to Eq.~\eqref{eq:precision}, the maximum precision in this case is   \begin{align}
    \mathcal{P}_{\ms{max}}^{\,\ms{F}} = \sqrt{\max_\phi\mathcal{F}_\phi} = 2\left(\max_\phi V_{|\psi\rangle}(\hat{X}_\phi)\right)^{1/2} .
\end{align} 
The corresponding metrological power for displacement sensing is 
\begin{align}
   \mathcal{M}_{\ket{\psi}}^{\,\ms{F}} = 4\left(\max_\phi V_{|\psi\rangle}(\hat{X}_\phi) - \frac{1}{2}\right).\label{eq:MPQ}
\end{align}  
The difficulty of calculating the maximum precision for all other transformations, in which $\hat G$ has terms involving products of two or more mode operators, is the complexity of the expression for the QFI when considering one or more input quantum states, additional coherent inputs, and an arbitrary passive linear network~\cite{Szczy16Multi, PhysRevA.94.042342,Ciampini:2016aa,PhysRevA.98.012114,Gessner18,Matsubara:2019aa,Liu:2019aa,Li:2020aa,ALBARELLI, PhysRevLett.111.130503, PhysRevLett.123.250503, YadinPRX18,PhysRevA.97.032329,GePRL2018, Kwon19}. We show in the section of Results that this expression can be made tractable by a judicious application of normal ordering~\cite{SZ}. It is then possible to maximize the result over all passive linear networks even when $\hat G$ involves arbitrarily high-order nonlinearities and multiple nonclassical input modes. 

\bibliography{unify-met-powers}

\section*{Author contributions statement}

W.G. and K.J. conceived this project and derived the results of the metrological power and the maximum precision. All authors wrote and reviewed the manuscript. 

\section*{Supplemental Materials}

\appendix
\section{Derivations of Phase Metrology with a Single-Mode Nonclassical State \label{AppenA}}
According to Eq.~(2) in the main text, $:\!\mathcal{F}_{jk}\!:$ now consists of sums of products of four quantities that are either the mode operator $\hat a_0$ (or its Hermitian conjugate) or one of the $f_j$ (or $f_j^*$). A simplification now results because i) the terms with less than two factors of $\hat a_0$ or $\hat a_0^\dagger$ cancel, ii) also vanishing are those terms containing two factors of the mode operators in which \textit{both} factors appear in either the first expectation value in the term $\langle \hat{n}_k \rangle\langle \hat{n}_j \rangle$ or the second, and iii) since $|\langle \hat a_0^n \rangle|$ and $\langle (\hat a_0^\dagger \hat a_0)^n \rangle$ are very much smaller than $f_j^n$ and $|f_j|^{2n}$, of the remaining terms only those with the fewest factors of the mode operator contribute appreciably. The result is that 
\begin{align}
  \langle \hat{n}_j \hat{n}_k \rangle & = \left[ (u_{k0}^* f_k) (u_{j0}^* f_j)^* + \mbox{c.c.} )\right] \langle  \hat a_0^\dagger \hat a_0  \rangle  + \left[ (u_{k0}^* f_k)^* (u_{j0}^* f_j)^*  \langle  a_0^2  \rangle + \mbox{H.c} \right]  = 2 |u_{k0} f_k| |u_{j0} f_j| \langle \, : \! \hat X_{\phi_k} \hat X_{\phi_j} \! : \, \rangle 
    \label{genform2}
\end{align}
where $\phi_k = \arg[u_{k0}^* f_k]$. Similarly $\langle \hat{n}_j \rangle = \sqrt{2} |u_{k0} f_k| \langle\hat X_{\phi_k} \rangle $.

Putting the expression for $:\!\mathcal{F}_{jk} \! :$ into Eq.~(3) in the main text, we can now factor the double summation to write the metrological advantage in terms of a single variance:
\begin{align} 
  \Delta \mathcal{P}^2_{\ket{\psi}}\left(N_{\text{c}},\hat{\mathbf{n}}, \mbox{\textbf{w}}\right) & \leq \frac{1}{|\mbox{\textbf{w}}|^4}\sum_{jk} w_j w_k :\!\mathcal{F}_{jk} \! : = \frac{8}{|\mbox{\textbf{w}}|^4}  \sum_{kj} : \!  C(  w_k |u_{k0} f_k|  \hat X_{\phi_k} ,  w_j |u_{j0} f_j| \hat X_{\phi_j} ) \! : \, \nonumber\\
   & = \frac{8}{|\mbox{\textbf{w}}|^4}  : \!  V\left( \sum_k  w_k |u_{k0} f_k|\hat X_{\phi_k}  \right)  \! : = \frac{8 |z|^2}{|\mbox{\textbf{w}}|^4}   : \!  V(\hat X_{\phi})\! :  \label{loopy}
\end{align} 
which proves Eq.~(6) in the main text.

To  maximize $|z|^2$ in Eq.(\ref{loopy}), we first note that since the $u_{jk}$ are the elements of a single unitary transformation, $U$, the vectors  
\begin{align}
    \bm{g} & \equiv  (f_0,\ldots,f_{m-1})/|\alpha| , \label{gdef} \\
    \bm{u} & = (u_{00}, \ldots, u_{(m-1)0}) , 
\end{align}  
both have unit norm. It is now useful to define the vector 
\begin{align}
    \bm{x} = (u_{00}^* f_0,\ldots,u_{(m-1)0}^* f_{m-1})/ |\alpha| . 
\end{align} 
It follows from the Cauchy-Schwartz inequality that the $l_1$ norm of $\bm{x}$, denoted by $||\bm{x}||_1$, is bounded by unity: 
\begin{align}
    ||\bm{x}||_1^2 = \left(\sum_j|u_j^{\ast}g_j|\right)^2 \leq |\bm{u}|^2 |\bm{g}|^2 = 1.
\end{align} 
With these definitions we can now write 
\begin{align} 
|z|^2=N_{\text{c}} \biggl| \sum_j w_j x_j\biggr|^2 \le N_{\text{c}} |\mbox{\textbf{w}}|^2 \left|\bm{x}\right|^2 \leq N_{\text{c}} |\mbox{\textbf{w}}|^4,  \label{ineqZ}
\end{align} 
Here the first inequality is the Cauchy-Schwartz inequality. The second inequality follows from the fact that the first inequality is saturated when $ \bm{x} = ||\bm{x}||_1\mbox{\textbf{w}}e^{i\varphi}$ for an arbitrary phase $\varphi$ and $||\bm{x}||_1 \leq 1$.

\subsection{Explicit networks achieving optimal precision for a single-mode nonclassical state}
\label{AppA}

In the distributed metrology scheme with a single-mode nonclassical state, the optimal precision, Eq.~(9) in the main text, is achieved when the inequality Eq.~\eqref{ineqZ} is saturated. Here we derive explicit solutions that do so. 

The inequality is saturated when $ \bm{x} = ||\bm{x}||_1\mbox{\textbf{w}}e^{i\varphi}$ for an arbitrary phase $\varphi$ and $||\bm{x}||_1 = 1$. We can rewrite $ \bm{x}$ as
\begin{align}
        \bm{x} = \frac{1}{|\alpha|} (|u_{00}| \tilde{f}_0,\ldots,|u_{(m-1)0}| \tilde{f}_{m-1}),
\end{align}
where $\tilde{f}_j=\sum_{k=0}^{m-1}u_{jk}e^{-i\varphi_j}\alpha_k$ and $\varphi_j=\arg[u_{j0}]$. The conditions $ \bm{x} = ||\bm{x}||_1\mbox{\textbf{w}}e^{i\varphi}$ and $||\bm{x}||_1 = 1$ now lead to
\begin{align}
    \frac{\tilde{f}_j}{|u_{j0}|}=\text{sgn}(w_j)|\alpha|e^{i\varphi},
    \label{eq:condition2-1}
\end{align}
and 
\begin{align}
    |u_{j0}|^2=|w_j|.
\end{align}
 Without loss of generality, we choose $\varphi$ such that $\arg[\alpha]=\varphi$. Restoring the phase $\varphi_j$ in $u_{j0}$, Eq.~\eqref{eq:condition2-1} can be rewritten as $\tilde{f}_j=u_{j0}\text{sgn}(w_j)\alpha$.  After some algebra we obtain 
\begin{align}
    \tilde{\bm{u}}_j\cdot\bm{t}=0, \;\; j=0,1,\cdots,m-1 , 
\end{align}
where 
\begin{align}
    \tilde{\bm{u}}_j \equiv (u_{j0},u_{j1},\cdots,u_{jm-1}) , 
\end{align}
is the $j^{\msi{th}}$ column of the unitary matrix $U$ and possible choices of $\bm{t}$ are 
\begin{align}
    \bm{t} = \left( \alpha_0, \alpha_1, \ldots, \alpha_{k} - \text{sgn}(w_j)\frac{u_{j0}}{u_{jk}}\alpha, \alpha_{k+1}, \ldots, \alpha_{m-1} \right)
\end{align} 
for some choice of $k$ from $0$ to $m-1$. Since the column vectors $\tilde{\bm{u}}_j$ form a complete orthonormal basis of the $m$ dimensional complex space, $\bm{v}$ must be zero to satisfy the orthogonality relations for all $j$. The general solution is  
\begin{align}
    \alpha_k = \delta_{lk}  \alpha e^{i\chi} , \;\;\; k = 0, 1, \ldots, m-1 , 
\end{align}
with $\chi = \arg[u_{j0}u^{\ast}_{jk}w_j]$ for all $j$. Thus we have to choose $u_{j0}$, $u_{jk}$ and $w_j$ such that $\chi$ is the same for all $j$.
%Note that since mode 1 is the only classical input mode with a nonzero amplitude, the number of input modes is $m=2$ and thus the index $j$ takes only the values $0$ and $1$. 
Choosing $k = 1$ and $|u_{j0}|=|u_{j1}|$, this solution is an MZI with the non-classical state input to one port and a coherent state with amplitude $\alpha e^{i\chi}$ input to the other.

\begin{figure*}[t]
%\leavevmode\includegraphics[width = .75 \columnwidth]{fig2.pdf}
\leavevmode\includegraphics[width = 1\columnwidth]{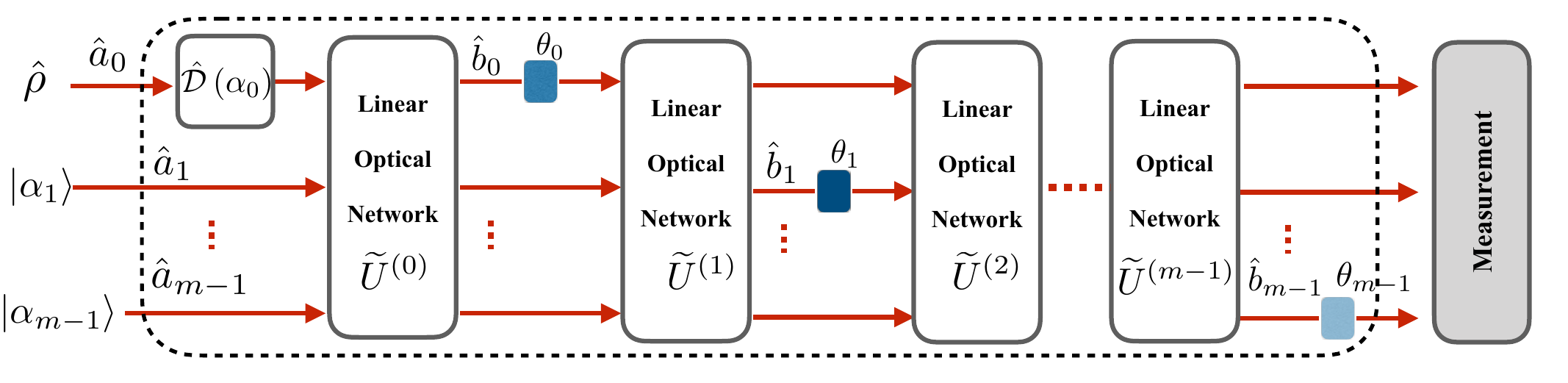}
\caption{(Color online) The most general linear metrology scheme for measuring $N$ single-mode quantities. As in the distributed metrology scheme of Fig.~1 in the main text, it consists of a passive linear network and $m$ input modes, one of which may contain a nonclassical state that is  displaced by a coherent amplitude $\alpha_0$. The general linear network differs from distributed metrology in that it is broken into $m$ segments which may be applied between the probe transformations (the phase shifts $\theta_j$).} 
\label{fig:scheme2} 
\end{figure*}

\section{Maximum precision for the general configuration of Fig.~\ref{fig:scheme2}
} 

%\begin{widetext}
%\end{widetext}
Here we would like to show that a general configuration of multi-mode quantum sensing scheme (Fig.~\ref{fig:scheme2}) does not provide more advantage in the precision than the distributed quantum sensing scheme in the main manuscript. 

The difference between the distributed metrology scheme analyzed in the main manuscript and the most general scheme here is that now each of the modes $\hat{b}_k$ is related to the input modes $\hat{a}_j$ by a different unitary,  
\begin{align}
   U_k  = \prod_{j=0}^k \widetilde{U}^{(j)}. 
\end{align}
Thus $\hat{b}_k = \sum_{j=0}^{m-1} u^{(k)}_{kj}\, a_j$ where $u^{(k)}_{kj}$ are the matrix elements of $U_k$. The analysis for this scheme is identical to that in the distributed metrology until we come to maximize the quantity $|z|^2$ in Eq.(\ref{loopy}), where now we have 
\begin{align}
    z & = \sum_k  w_k u^{(k)}_{k0} f_k^* 
\end{align} 
and $f_k = \sum_{j} u^{(k)}_{kj}\alpha_j$. There is considerably more freedom in choosing the elements $u^{(k)}_{k0}, k=1,\ldots, m$: since each comes from a different unitary they no longer need to form a unit vector.  

To maximize $z$ we first note that since 
\begin{align}
    |f_j|^2 & = \Biggl| \sum_{k=0}^{m-1} u_{jk}^{(j)} \alpha_k \Biggr|^2 
    \le \Biggl(\,\sum_{k=0}^{m-1}\bigl|u_{jk}^{(j)}\bigr|^2 \Biggr)  \,\Biggl(\,\sum_{l=0}^{m-1}|\alpha_l|^2 \Biggr) = N_{\text{c}} , 
\end{align}
we have 
\begin{align}
    |z|^2 &  \leq  N_{\text{c}} \, \Biggl(\, \sum_j \bigl| w_j u_{j0}^{(j)} \bigr|\Biggr)^2 \leq N_{\text{c}} \, \biggl( \sum_j w_j^2 \biggr) \, \biggl(\sum_k |u_{k0}^{(k)}|^2 \biggr) .  
\end{align} 
The last inequality is saturated when $u_{j0}^{(j)} = c w_j$ for every $j$ for some constant $c$. Since  $|u_{j0}^{(j)}| \leq 1$ we must have $|c| \leq 1/w_{\ms{max}}$ where $w_{\ms{max}} \equiv \max_j |w_j|$. So we can now write  
\begin{align} 
   |z|^2  & \leq N_{\text{c}} \, \biggl( \sum_j w_j^2 \biggr)^2 \, |c|^2 \leq \frac{N_{\text{c}}}{w_{\ms{max}}^2}  \biggl( \sum_j w_j^2 \biggr)^2 = N_{\text{c}} \frac{|\mbox{\textbf{w}}|^4}{w_{\ms{max}}^2} . 
\end{align}
We can saturate this bound while selecting any value for $\phi$ merely by choosing the phases of $\alpha_j$ and $u_{jk}^{(j)}$.

Substituting the above maximum in the Eq.(\ref{loopy}), we can now write down the nonclassical contribution to the precision for a general phase measurement with a given weighting distribution $\{w_i\}$, which is  
\begin{align}
  \Delta \mathcal{P}^2_{\ket{\psi}}\left(N_{\text{c}},\hat{\mathbf{n}}, \mbox{\textbf{w}}\right) = \frac{8 N_{\text{c}}}{w_{\ms{max}}^2} \max_\phi :\!V_{\ket{\psi}}(\hat{X}_{\phi})\!: \; = \frac{2N_{\text{c}}}{w_{\ms{max}}^2}  \mathcal{M}_{|\psi\rangle}^{\ms{\,F}}.
  \label{loop2}
\end{align}

To obtain the maximum possible precision of $m$ phase shifts, we must choose the weights $\mbox{\textbf{w}}$ to give the minimal value for $w_{\ms{max}} = \max_i |w_i|$. This is achieved by setting $w_i = 1/m$ for all $i$ so that the measured quantity is simply the average of all the phase shifts. The result is  
\begin{align} 
  \max_{\mbox{\textbf{w}}}\Delta \mathcal{P}^2_{\ket{\psi}}\left(N_{\text{c}},\hat{\mathbf{n}}, \mbox{\textbf{w}}\right)  =  2m^2N_{\text{c}} \mathcal{M}_{|\psi\rangle}^{\ms{\,F}} . 
  \label{loop3}
\end{align} 
As compared to distributed metrology (Eq.~(9) in the main text), there is an additional factor $m^2$ in the nonclassical contribution. Examining the conditions required to saturate the upper bounds we find that the optimal scheme requires applying all the phase shifts in sequence to the same mode, something that is precluded in distributed metrology. Because of this the factor of $m^2$ is a purely \emph{classical} effect: it is due to the fact that we are estimating the average of $m$ quantities, where the \textit{sum} of those quantities is applied to a single mode and thus measured directly. Compared to applying just one of the quantities, the resulting error is therefore reduced by a factor of $m$.

\section{Metrology with an arbitrary transformation, a single-mode quantum state, and arbitrary linear resources} 

\subsection{System setup}
We want to generalize the phase-shifting parameter-encoding scheme to an arbitrary single-mode transformation  $\hat{G}(\hat{b},\hat{b}^\dagger)$  for metrology at the probe, where $\hat{G}$ is given by a sum of products of $\hat{b}$ and $\hat{b}^\dagger$, up to and including a number of products that contain $p$ of these operators (but no more than $p$). First we rearrange the operators in every product by using $[\hat{b},\hat{b}^\dagger]=1$ so that $\hat{G}$ is written in the normally-ordered form as $\hat{G} = \sum_{q=0}^p\sum_{j=0}^{q}\kappa_{qj} \hat{b}^{\dagger(q-j)}\hat{b}^j$. 

We will find that terms with $q$ contribute a value proportional to $N_{\text{c}}^{q-1}$. Thus only leading contributing terms will be those of order $p$ for large $|\alpha|$. Thus for the purpose of calculating the metrological power, we can write all Hermitian operators $\hat{G}$ as 
\begin{align}
    \hat{G} =  \sum_{j=0}^p \kappa_j \hat{C}_j
\end{align}
and define
\begin{align}
    \hat{C}_j = \hat{b}^{\dagger (p-j)} \hat{b}^{j}.
    \label{eq:Cj}
\end{align}
Since $\hat{G}$ is an observable, it is Hermitian by requiring $\kappa_j=\kappa_{p-j}^{\ast}$.
As per the main text on the distributed metrology,
\begin{align}
    \mathcal{M}_{\ket{\psi}}\left(N_c,\hat{\mathbf{G}}, \mbox{\textbf{w}}\right)=\max_{\text{PLN}}\lim_{N_{\text{c}}/N_{\text{q}} \rightarrow\infty}\frac{1}{|\mbox{\textbf{w}}|^4}\sum_{uv} w_{u} w_{v} :\!\mathcal{F}_{uv} \! :,\label{eq:Mgen}
\end{align}
where
\begin{align}
   :\! \mathcal{F}_{uv} \!: & = 4 \biggl[  \biggl\langle :\! \hat{G}_u \hat{G}_v\!: \biggr\rangle - \biggl\langle \hat{ G}_u   \biggr\rangle  \biggl\langle  \hat{G}_v  \biggr\rangle \biggr]=4\sum_{j,k}\kappa_j \kappa_k \left[ \langle :\! \hat{C}_j^{(u)} \hat{C}_k^{(v)} \!: \rangle - \langle \hat{C}_j^{(u)} \rangle \langle \hat{C}_k^{(v)} \rangle \right],
\end{align} 
and $\hat{G}_u$ and $\hat{C}_j^{(u)}$ are defined by replacing $\hat{b}$ ($\hat{b}^{\dagger}$) with $\hat{b}_u$ ($\hat{b}_u^{\dagger}$) in $\hat{G}$ and $\hat{C}_j$, respectively.

\subsection{Evaluation of $:\! \mathcal{F}_{uv} \!:$}

We now replace 
\begin{align}
    \hat{b}_u \rightarrow u_{u0}\hat{a}_0 + f_u
\end{align}
in $\langle :\! \hat{C}_j^{(u)} \hat{C}_k^{(v)} \!: \rangle - \langle \hat{C}_j^{(u)} \rangle \langle \hat{C}_k^{(v)} \rangle$, where    $f_{u} = \sum_{l=0}^{m-1} u_{ul} \alpha_{l}$ and  $u_{ul}$ are the elements of the $u^{th}$ row of the unitary matrix $U$ in the distributed quantum metrology scheme (Fig.~1 in the main text). We obtain 
\begin{align}
    \langle (u_{u0}\hat{a}_0 + f_u)^{\dagger s}(u_{v0}\hat{a}_0 + f_v)^{\dagger t}(u_{u0}\hat{a}_0 + f_u)^{j}(u_{v0}\hat{a}_0 + f_v)^{k} \rangle - \langle (u_{u0}\hat{a}_0 + f_u)^{\dagger s}(u_{u0}\hat{a}_0 + f_u)^j \rangle \langle  (u_{v0}\hat{a}_0 + f_v)^{\dagger t}(u_{v0}\hat{a}_0 + f_v)^k \rangle 
    \label{genform}
\end{align}
where $s=p-j$ and $t=p-k$. We now recall that for large $|\alpha|$, the dominant terms that appear from expanding the above expression have exactly two factors of the operators (either $\hat{a}_0$ or $\hat{a}_0^\dagger$) where one factor must come from the first expectation value in the second term above and the other from the second expectation value. Counting the dominant terms, we have 
\begin{align} 
\lim_{N_{\text{c}}/N_{\text{q}} \rightarrow\infty} \langle :\! \hat{C}_j^{(u)} \hat{C}_k^{(v)} \!: \rangle - \langle \hat{C}_j^{(u)} \rangle \langle \hat{C}_k^{(v)} \rangle &= jkf_u^{\ast s}f_v^{\ast t}f_u^{j-1}f_v^{k-1}u_{u0}u_{v0}\left(\braket{\hat{a}_0^2}-\braket{\hat{a}_0}^2\right) +stf_u^{\ast s-1}f_v^{\ast t-1}f_u^{j}f_v^{k}u^{\ast}_{u0}u^{\ast}_{v0}\left(\braket{\hat{a}_0^{\dagger 2}}-\braket{\hat{a}_0}^{\dagger 2}\right)\nonumber\\
&+jtf_u^{\ast s}f_v^{\ast t-1}f_u^{j-1}f_v^{k}u_{u0}u^{\ast}_{v0}\left(\braket{\hat{a}_0^{\dagger}\hat{a}_0}-|\braket{\hat{a}_0}|^2\right) +ksf_u^{\ast s-1}f_v^{\ast t}f_u^{j}f_v^{k-1}u^{\ast}_{u0}u_{v0}\left(\braket{\hat{a}_0^{\dagger}\hat{a}_0}-|\braket{\hat{a}_0}|^2\right)\nonumber\\
&=f_u^{\ast s-1}f_v^{\ast t-1}f_u^{j-1}f_v^{k-1} \left[\left\langle:\! \left(sf_uu_{u0}^{\ast}\hat{a}_0^{\dagger}+jf_u^{\ast}u_{u0}\hat{a}_0\right)\left(tf_vu_{v0}^{\ast}\hat{a}_0^{\dagger}+kf_v^{\ast}u_{v0}\hat{a}_0\right)\!:\right\rangle\right.\nonumber\\
&\left.-\left\langle sf_uu_{u0}^{\ast}\hat{a}_0^{\dagger}+jf_u^{\ast}u_{u0}\hat{a}_0\right\rangle\left\langle tf_vu_{v0}^{\ast}\hat{a}_0^{\dagger}+kf_v^{\ast}u_{v0}\hat{a}_0 \right\rangle  \right].
 \label{Gen}
\end{align}

Thus, we find 

\begin{align}
    \lim_{N_{\text{c}}/N_{\text{q}} \rightarrow\infty}:\! \mathcal{F}_{uv} \!:=4\left(\braket{:\!\hat{A}_u\hat{A}_v\!:}-\braket{\hat A_u}\braket{\hat{A}_v}\right),\label{eq:Fuv}
\end{align}
where $\hat{A}_u=\sum_{j=0}^p\kappa_jf_u^{\ast p-j-1}f_u^{j-1}\left((p-j)f_uu_{u0}^{\ast}\hat{a}_0^{\dagger}+jf_u^{\ast}u_{u0}\hat{a}_0\right)$.

\subsection{Unification of metrological powers}
By substituting $:\! \mathcal{F}_{uv} \!:$ in Eq.~\eqref{eq:Mgen}, we obtain
\begin{align}
    \mathcal{M}_{\ket{\psi}}\left(N_c,\hat{\mathbf{G}} ,\mbox{\textbf{w}}\right)=\max_{\text{PLN}}\frac{4}{|\mbox{\textbf{w}}|^4}\sum_{uv} w_{u} w_{v} \left(\braket{:\!\hat{A}_u\hat{A}_v\!:}-\braket{\hat{A}_u}\braket{\hat{A}_v}\right)=\max_{\text{PLN}}\frac{8 |z|^2}{|\mbox{\textbf{w}}|^4} \!:\!V_{\ket{\psi}}( \hat{X}_{\phi})\!: ,
\end{align}
where $\phi=\arg z$ and
\begin{align}
    z=\sum_uw_uu_{u0}^{\ast}\sum_{j=0}^p\kappa_jf_u^{\ast p-j-1}f_u^{j}(p-j).
\end{align}
Here we note that (i) the relation between the general metrological power and the quadrature variance is guaranteed by the fact that $\hat{A}_u$ is a quadrature operator of $\hat{a}_0$ since $\sum_{j=0}^p\kappa_jf_u^{\ast p-j-1}f_u^{j}(p-j)u_{u0}^{\ast}=\left(\sum_{j=0}^p\kappa_jf_u^{\ast p-j}f_u^{j-1}ju_{u0}\right)^{\ast}$ using $\kappa_j=\kappa_{p-j}^{\ast}$. (ii) the expression of $z$ reduces to that of distributed quantum metrology when we take $p=2$ and $\kappa_j=\delta_{j1}$, where $\delta_{jk}$ is the Kronecker delta.

To obtain the connection between the general metrological power and that of the displacement sensing, $\mathcal{M}_{\ket{\psi}}^{\,\ms{F}} = 4\max_\phi :\!V_{|\psi\rangle}(\hat{X}_\phi)\!:$, we maximize the $:\!V_{\ket{\psi}}( \hat{X}_{\phi})\!:$ through the phase of $z$ and the amplitude $\frac{8 |z|^2}{|\mbox{\textbf{w}}|^4}$ using an arbitrary linear network. We find the maximum amplitude depends on $\kappa_j$ of the observable $\hat{G}$, thus for any pure state $\ket{\psi}$, we have 

\begin{align}
    \mathcal{M}_{\ket{\psi}}\left(N_c,\hat{\mathbf{G}}\right)=\left(\max_{\text{PLN}, \phi=\phi_m}\frac{2 |z|^2}{|\mbox{\textbf{w}}|^4}\right)\mathcal{M}_{\ket{\psi}}^{\ms{F}},
\end{align}
where $\phi_m$ is the phase when $:\!V_{\ket{\psi}}( \hat{X}_{\phi})\!:$ is maximized for the state $\ket{\psi}$, and $\phi=\phi_m$ is the constraint to make sure $\mathcal{M}_{\ket{\psi}}\left((N_c,\hat{\mathbf{G}}\right)\propto\mathcal{M}_{\ket{\psi}}^{\ms{F}}$.

To maximize the amplitude, we consider
\begin{align}
    |z|^2\leq\left|\sum_uw_uu_{u0}^{\ast}\right|^2\left|\sum_{j=0}^p\kappa_j(p-j)f_u^{\ast p-j-1}f_u^{j}\right|^2 = \left|\sum_uw_uu_{u0}^{\ast} |f_u|^{p-1} \right|^2 \left|\sum_{j=0}^p\kappa_{p-j}je^{i2(p-j)\gamma}\right|^2 
    \le \sum_u |f_u|^{2(p-1)}|\mbox{\textbf{w}}|^4\max_{\gamma}\left|\sum_{j=0}^p\kappa_{j}je^{i2j\gamma}\right|^2,
\end{align}
where the second inequality is obtained according to the results in the analysis of the distributed  phase sensing in Appendix~\ref{AppenA} and the relation $|f_u|\le|\alpha|$. The maximal value is achieved when $f_u$ is nonzero for only one element. Details can be found below Eq.~(31) in the main text. The phase $\gamma\equiv\arg f_u$ is assumed to be the same for different $u$ and $|z|$ is maximized by choosing $\gamma$ that maximize the coefficient $B(\bm{\kappa})\equiv\max_{\gamma}\left|\sum_{j=0}^p\kappa_{j}je^{i2j\gamma}\right|^2$. The optimal phase in the quadrature variance can be obtained by choosing the phase of $u_{u0}$. Now we have
\begin{align}
    \mathcal{M}_{\ket{\psi}}\left(N_c,\hat{\mathbf{G}}\right)=2N_{\text{c}}^{p-1}B(\bm{\kappa})\mathcal{M}_{\ket{\psi}}^{\ms{F}}.
\end{align}
Similar to the distributed quantum sensing scheme, the above relation also extends to mixed states according to the convex roof of quantum Fisher information. Thus we have proved that the metrological powers of a single-mode quantum state for any unitary transformations are all proportional to that of displacement sensing.

\subsection{Some examples}
Here we consider a special class of observables where $\kappa_j$ are all real. Using $\kappa_j=\kappa_{p-j}$, we have
\begin{align}
    \max_{\gamma}\left|\sum_{j=0}^p\kappa_{j}je^{i2j\gamma}\right|^2=\frac{p}{2}\sum_{j=0}^p\kappa_j
\end{align}
We consider that $\sum_{j=0}^p\kappa_j=1$ as a normalization condition in accordance with the usual phase sensing observable, where $\hat{G}=\frac{1}{2}\left(\hat{b}^{\dagger}\hat{b}+h.c.\right)$, So the general metrological power becomes
\begin{align}
     \mathcal{M}_{\ket{\psi}}\left((N_c,\hat{\mathbf{G}}\right)=\frac{p^2}{2}N_{\text{c}}^{p-1}\mathcal{M}_{\ket{\psi}}^{\ms{F}}
\end{align}

Now we consider some examples. For a phase measurement we have 
\begin{align}
    \hat G = \hat b^\dagger \hat b = \frac{1}{2} \hat b^\dagger\hat b + \mbox{H.c.}
\end{align}
so that only one of the $\kappa$'s is non-zero, being $\kappa_1 = 1$, and $p=2$. So we have 
\begin{align}
    \mathcal{M}_{\ket{\psi}}\left((N_c,\hat{\mathbf{G}}\right) & =  2 N_{\text{c}}  \mathcal{M}_{\ket{\psi}}^{\ms{F}} 
\end{align}
in agreement with our previous result. For a Kerr non-linearity we have 
\begin{align}
    \hat G = \hat b^{\dagger 2} \hat b^2 = \frac{1}{2} \hat b^{\dagger 2} \hat b^2 + \mbox{H.c.}
\end{align}
with $\kappa_2 = 1$ and $p=4$, so that 
\begin{align}
   \mathcal{M}_{\ket{\psi}}\left((N_c,\hat{\mathbf{G}}\right) & =  8 N_{\text{c}}^{3}  \mathcal{M}_{\ket{\psi}}^{\ms{F}}
\end{align}

\subsection{The maximum precision in terms of total photon number} 

The maximum precision for measuring a single phase shift, as per Eq.~(12) in the main text , is  
\begin{align}
     \textstyle  \mathcal{P}_{|\psi\rangle} (N_c, \hat{n}) = \sqrt{8[\max_\phi V_{|\psi\rangle}(\hat{X}_\phi) ] N_{\ms{c}}} .
     \label{dwond}
\end{align}
The state that has the maximum quadrature variance, $\max_\phi V_{|\psi\rangle}(\hat{X}_\phi)$, for a given energy is a squeezed vacuum state \cite{ge2020operational}. This maximum variance is given by 
\begin{align}
    \max_\phi V_{|\psi\rangle}(\hat{X}_\phi) = \frac{1}{2}e^{2r} 
\end{align}
where $r$ is called the squeezing parameter and characterizes the state~\cite{SCHUMAKER1986317}. The energy (the average number of photons) of the state is given by 
\begin{align}
    M = \sinh^2(r) = \frac{1}{4}\left( e^{2r} + e^{-2r} - 2 \right) \equiv  \frac{1}{4}\left( K + \frac{1}{K} - 2 \right)
\end{align}
where we have defined $K =  e^{2r}$. Now solving for $K$, we find
\begin{align}
  K & =  (1+2M) + \sqrt{(1+2M)^2 - 1}  =  (1+2M) + \sqrt{4M(M+1)} > 4M
\end{align} 
Substituting into Eq.(\ref{dwond}) the maximum precision is 
\begin{align}
    \sqrt{4NK} > \sqrt{16 N M} = 4 \sqrt{N M}
\end{align}
We note that the maximum precision for an MZI with a squeezed vacuum is~\cite{Pezze08} 
\begin{align}
    \sqrt{NK + M} \approx \sqrt{NK}  > \sqrt{4NM}.
\end{align}
The factor $2$ difference of the asymptotic precision in the limit of $N_c\rightarrow \infty$ can be resolved with a unified definition of the global parameter to be estimated (see Eq.~(44) in the main text).

\section{Metrology with an arbitrary transformation, a multi-mode quantum state, and arbitrary linear resources}

\subsection{Classical precision}

We have determined that the quantum contribution of the precision for a general sensing scheme from a single-mode quantum state. To obtain the total precision, we need the classical contribution, which means whatever terms are generated when we normally order the products of $C_j$ in Eq.~\eqref{eq:Cj}. In the limit of $|\alpha|\rightarrow \infty$, we keep only the leading order of $\hat b^{\dagger m}\hat b^{n}$. 
\begin{align} 
    \mathcal{F}_{uu} & = 4 \sum_{j,k}\kappa_j \kappa_k  \left\langle \hat{C}_j\hat{C}_k \right\rangle  = 4 \sum_{j,k}\kappa_j \kappa_k  \left\langle \hat{b}^{\dagger (p-j)} \hat{b}^{j} \hat{b}^{\dagger (p-k)} \hat{b}^{k}  \right\rangle \\
   & = 4 \sum_{j,k}\kappa_j \kappa_k  \left\langle \hat{b}^{\dagger (p-j)} \left[\hat{b}^{\dagger (p-k)} \hat{b}^{j} + (p-k) j \hat{b}^{\dagger (p-k-1)} \hat{b}^{j-1} \right] \hat{b}^{k}  \right\rangle \\
   & = 4 \sum_{j,k}\kappa_j \kappa_k  \left\langle \hat{b}^{\dagger (p-j)} \hat{b}^{\dagger (p-k)} \hat{b}^{j} \hat{b}^{k}  \right\rangle + \kappa_j \kappa_k (p-k) j \left\langle \hat{b}^{\dagger (p-j)} \hat{b}^{\dagger (p-k-1)} \hat{b}^{j-1} \hat{b}^{k}  \right\rangle \\
   & = 4 \sum_{j,k=0}^p\kappa_j \kappa_k  \left\langle :\!  \hat{C}_j\hat{C}_k \! : \right\rangle  + 4 \sum_{j,k=1}^{p-1} \kappa_j \kappa_k (p-k) j  f_{u}^{* (2p-j-k-1)} f_{u}^{j+k-1}  \\
   & = 4 \sum_{j,k=0}^p\kappa_j \kappa_k  \left\langle :\!  \hat{C}_j\hat{C}_k \! : \right\rangle  + 4 \left|f_{u} \right|^{2p-2} \sum_{j,k=1}^{p-1} \kappa_j \kappa_k (p-k) j  e^{-2i (p-j-k)\gamma_{u}}  \\
   & = 4 \sum_{j,k=0}^p\kappa_j \kappa_k  \left\langle :\!  \hat{C}_j\hat{C}_k \! : \right\rangle  + 4 \left|f_{u} \right|^{2p-2} \biggl( \sum_{k=1}^{p-1} \kappa_k (p-k) e^{-2i (p-k)\gamma_{u}} \biggr) \biggl( \sum_{j=1}^{p-1} \kappa_j  j  e^{2ij\gamma_{u}} \biggr) \\
   & = 4 \sum_{j,k=0}^p\kappa_j \kappa_k  \left\langle :\!  \hat{C}_j\hat{C}_k \! : \right\rangle  + 4 \left|f_{u} \right|^{2p-2} \biggl( \sum_{l=0}^{p-1} \kappa_{(p-l)} l e^{-2i l\gamma_{u}} \biggr) \biggl( \sum_{j=1}^{p-1} \kappa_j  j  e^{2ij\gamma_{u}} \biggr) \\
   & = 4 \sum_{j,k=0}^p\kappa_j \kappa_k  \left\langle :\!  \hat{C}_j\hat{C}_k \! : \right\rangle  + 4 \left|f_{u} \right|^{2p-2} \Biggl|  \sum_{k=1}^{p-1} \kappa_k  k  e^{2ik\gamma_{u}} \Biggr|^2 
\end{align} 
Here we have used the fact that, for $k \leq n$, 
\begin{align}
    \hat b^k \hat b^{\dagger n}  & =   \hat b^{\dagger n} \hat b^k  +   k n  \hat b^{\dagger n-1} \hat b^{k-1}   +   \frac{k (k-1) n (n-1)}{2!}  \hat b^{\dagger n-2}\hat b^{k-2} + \cdots + \frac{n (n-1) \cdots (n-k+1) k (k-1) \cdots (k-k+1)}{k!}  \hat b^{\dagger n-k} \\
    & =  \hat b^{\dagger n}  \hat b^k  +   \frac{k! n!}{(k-1)!(n-1)!1!}  \hat b^{\dagger n-1} \hat b^{k-1}   +  \frac{k! n!}{(k-2)!(n-2)!2!}  \hat b^{\dagger n-2}\hat b^{k-2} + \cdots + \frac{k! n!}{(k-k)!(n-k)!k!}  \hat b^{\dagger n-k} \\
    & =  \hat b^{\dagger n} \hat b^k + \sum_{j=1}^k \frac{k! n!}{(k-j)!(n-j)!j!}  \hat b^{\dagger n-j}\hat b^{k-j}
\end{align}
and we can include $n > k$ simply by extending this as 
\begin{align}
    \hat b^k  \hat b^{\dagger n}  & =   \hat b^{\dagger n} \hat b^k + \sum_{j=1}^{\min(k,n)} \frac{k! n!}{(k-j)!(n-j)!j!}  \hat b^{\dagger n-j}\hat b^{k-j}
\end{align}
The total classical QFI is the QFI when any quantum inputs are the vacuum. We have 
\begin{align} 
    \mathcal{P}^2_{\ms{c}}\left(N_{\text{c}},\hat n, \mbox{\textbf{w}}\right) & =  \frac{1}{|\mbox{\textbf{w}}|^4}\sum_{jk} w_j w_k :\!\mathcal{F}_{jk} \! : + \frac{4}{|\mbox{\textbf{w}}|^4}\sum_{u} w_u^2 \left|f_{u} \right|^{2p-2} \Biggl|  \sum_{k=1}^{p-1} \kappa_k  k  e^{2ik\gamma_{u}} \Biggr|^2  
\end{align} 
Since there are no quantum inputs the normally-ordered part is zero and so 
\begin{align}
      \mathcal{P}^2_{\ms{c}}\left(N_{\text{c}},\hat{\mathbf{G}}, \mbox{\textbf{w}}\right) & =   \frac{4}{|\mbox{\textbf{w}}|^4}\sum_{u} w_u^2 \left|f_{u} \right|^{2p-2} \Biggl| \sum_{k=1}^{p-1} \kappa_k  k  e^{2ik\gamma_{u}} \Biggr|^2 \leq \frac{4}{|\mbox{\textbf{w}}|^4} \sqrt{\sum_{u} w_u^4} \sqrt{\sum_u\left|f_{u} \right|^{4p-4}} \; B(\bm{\kappa})
\end{align}
To maximize this we need to set $|f_u|^{2(p-1)} = c w_u^2$. But we also have $\sum_u |f_u|^2 = N_{\ms{c}}$. This means that $\sum_u |f_u|^{2(p-1)} \leq N_{\ms{c}}^{p-1}$.If we set $w_u = 1/m$ we have $|f_u|^{2(p-1)} = N_{\ms{c}}^{p-1}/m^{p-1}$ and thus 
\begin{align}
  \mathcal{P}^2_{\ms{c}}\left(N_{\text{c}},\hat{\mathbf{G}}, \mbox{\textbf{w}}\right) & le \frac{4 N_{\ms{c}}^{p-1}}{m^{p-2}} B(\bm{\kappa})\le 4 N_{\ms{c}}^{p-1} B(\bm{\kappa}),
\end{align}
where the last inequality holds when $m=1$.

\subsection{Quantum contribution}
We now consider the precision when $q$ input modes have a joint non-classical state. In this case everything is the same as before except that the instead of 
\begin{align}
      \hat{b}_u \rightarrow u_{u0}\hat{a}_0 + f_u 
\end{align}
we now have 
\begin{align}
    \hat{b}_u \rightarrow \sum_{l=0}^{q-1} u_{ul}\hat{a}_l + f_u
\end{align}
were $q$ is the number of input modes carrying non-classical states. Recall that $f_{u} = \sum_{l=0}^{m-1} u_{ul} \alpha_{l}$. We can rewrite this sum over modes $a_l$ as a single mode operator: 
\begin{align}
  \sum_{l=0}^{q-1} u_{ul}\hat{a}_l = g_u \hat{d}_u 
\end{align} 
with   $g_u= \left( \sum_{l=0}^{q-1} |u_{ul}|^2 \right)^{\! 1/2}$  and 
  $d_u= \frac{1}{g_u} \sum_{l=0}^{q-1} u_{ul} a_l$. So we now have 
\begin{align}
    \hat{b}_u \rightarrow g_u \hat{d}_u + f_u
\end{align}
Following this modification through the analysis above, we have 
\begin{align}
    \lim_{N_{\text{c}}/N_{\text{q}} \rightarrow\infty}:\! \mathcal{F}_{uv} \!:=4\left(\braket{:\!\hat{A}_u\hat{A}_v\!:}-\braket{\hat A_u}\braket{\hat{A}_v}\right),\label{eq:Fuv}
\end{align}
where now 
\begin{align}
    \hat{A}_u=\sum_{j=0}^p\kappa_j f_u^{\ast p-j-1}f_u^{j-1}\left((p-j)f_u g_u d_u^\dagger +j f_u^{\ast} g_u d_u\right).
\end{align}
By substituting $:\! \mathcal{F}_{uv} \!:$ in Eq.~\eqref{eq:Mgen}, we obtain
\begin{align}
    \mathcal{M}_{\ket{\psi}}\left(N_c,\hat{\mathbf{G}},\mbox{\textbf{w}}\right)=\max_{\text{PLN}}\frac{4}{|\mbox{\textbf{w}}|^4}\sum_{uv} w_{u} w_{v} \left(\braket{:\!\hat{A}_u\hat{A}_v\!:}-\braket{\hat{A}_u}\braket{\hat{A}_v}\right)=\max_{\text{PLN}}\frac{8}{|\mbox{\textbf{w}}|^4} \!:\!V_{\ket{\psi}}\left( \sum_u w_u \hat{A}_u \right)\!: ,
\end{align}
where 
\begin{align}
    \sum_u w_u \hat{A}_u & = \sum_u w_u \sum_{j=0}^p\kappa_j f_u^{\ast p-j-1}f_u^{j-1}\left((p-j)f_u g_u^{\ast} d_u^\dagger +j f_u^{\ast} g_u d_u\right) \\
    & = \left[ \sum_{j=0}^{p-1}(p-j)\kappa_j e^{-i\gamma(p-2j-1)}\right]  \sum_u w_u g_u |f_u|^{p-1} \hat{d}_u + \mbox{H.c.}  \\
    & = \left[ \sum_{j=0}^{p-1}(p-j)\kappa_j e^{2ij\gamma}\right]  \sum_{l=0}^{q-1} \biggl(\sum_u w_u  |f_u|^{p-1}  u_{ul} \biggr) \hat{a}_l + \mbox{H.c.} \\
    & = \left| \sum_{j=0}^{p-1}j\kappa_j^* e^{2i(p-j)\gamma}\right|  \sqrt{\sum_{l=0}^{q-1} \biggl|\sum_u w_u  |f_u|^{p-1}  u_{ul} \biggr|^2} \hat{d} + \mbox{H.c.} \\
    & =   \left| \sum_{j=0}^{p-1} j\kappa_j e^{2ij\gamma}\right|  \sqrt{ \sum_{l=0}^{q-1} \biggl|\sum_u w_u  |f_u|^{p-1}  u_{ul} \biggr|^2} \hat{d} + \mbox{H.c.}
\end{align}
so 
\begin{align}
     \mathcal{M}_{\ket{\psi}}\left(N_c,\hat{\mathbf{G}},\mbox{\textbf{w}}\right) = \max_{\text{PLN}}\frac{8|z|^2}{|\mbox{\textbf{w}}|^4} \!:\!V_{\ket{\psi}}\left( \hat{X}_d \right)\!: 
     \label{SuppMpsi}
\end{align}
with
\begin{align}
    |z|^2 & =   \left| \sum_{j=0}^{p-1} j\kappa_j e^{2ij\gamma}\right|^2  \sum_{l=0}^{q-1} \biggl|\sum_u w_u  |f_u|^{p-1}  u_{ul} \biggr|^2 
\end{align}
and 
\begin{align}
    \hat d & = \frac{1}{\sqrt{\sum_{l=0}^{q-1} \biggl|\sum_u w_u  |f_u|^{p-1}  u_{ul} \biggr|^2}} \sum_{l=0}^{q-1} \biggl( \sum_u w_u  |f_u|^{p-1}  u_{ul} \biggr) \hat a_l 
\end{align}
Following the procedure in the main manuscript, we can maximize $|z|^2$ to be $N_c ^{p-1}B(\bm{\kappa})|\mbox{\textbf{w}}|^4$ and thus obtain the maximum nonclassical contribution to the precision.

\end{document}